\PassOptionsToPackage{pdfpagelabels=false}{hyperref}
\documentclass[fleqn,usenatbib]{mnras}

\usepackage{newtxtext,newtxmath}
\usepackage[T1]{fontenc}

\DeclareRobustCommand{\VAN}[3]{#2}
\let\VANthebibliography\thebibliography
\def\thebibliography{\DeclareRobustCommand{\VAN}[3]{##3}\VANthebibliography}

\usepackage{graphicx}
\usepackage{amsmath}
\usepackage{color}
\usepackage[dvipsnames]{xcolor}
\usepackage[title]{appendix}
\usepackage{orcidlink}
\usepackage{caption}
\usepackage{subcaption}

\input{macros.sty}
\input{hyperlink-year-only-natbib-patch}

\newcommand{\citepinprep}[1]{(#1 et al., \textcolor{blue}{in prep.})}
\newcommand{\citetinprep}[1]{#1 et al. (\textcolor{blue}{in prep.})}

\title[Merger Responses]{Merger Response of Halo Anisotropy Properties}

\author[K. Wang et al.]{
Kuan Wang\orcidlink{0000-0001-7690-2260}$^{1,2}$\thanks{E-mail: kuanwang@umich.edu},
Philip Mansfield\orcidlink{0000-0001-9863-5394}$^{3,4}$,
Dhayaa Anbajagane$^{5,6}$\orcidlink{0000-0003-3312-909X},
Camille Avestruz\orcidlink{0000-0001-8868-0810}$^{1,2}$\\
$^{1}$Department of Physics, The University of Michigan, Ann Arbor, MI 48109, USA\\
$^{2}$Leinweber Center for Theoretical Physics, University of Michigan, 450 Church St, Ann Arbor, MI 48109, USA\\
$^{3}$Kavli Institute for Particle Astrophysics \& Cosmology, P. O. Box 2450, Stanford University, Stanford, CA 94305, USA\\
$^{4}$SLAC National Accelerator Laboratory, Menlo Park, CA 94025, USA\\
$^{5}$Department of Astronomy and Astrophysics, University of Chicago, Chicago, IL 60637, USA\\
$^{6}$Kavli Institute for Cosmological Physics, University of Chicago, Chicago, IL 60637, USA
}

\date{Accepted XXX. Received YYY; in original form ZZZ}

\pubyear{2023}

\begin{document}
\label{firstpage}
\pagerange{\pageref{firstpage}--\pageref{lastpage}}
\maketitle

\begin{abstract}

Anisotropy properties --- halo spin, shape, position offset, velocity offset, and orientation --- are an important family of dark matter halo properties that indicate the level of directional variation of the internal structures of haloes.
These properties reflect the dynamical state of haloes, which in turn depends on the mass assembly history.
In this work, we study the evolution of anisotropy properties in response to merger activity using the IllustrisTNG simulations.
We find that the response trajectories of the anisotropy properties significantly deviate from secular evolution. 
These trajectories have the same qualitative features and timescales across a wide range of merger and host properties.
We propose explanations for the behaviour of these properties and connect their evolution to the relevant stages of merger dynamics.
We measure the relevant dynamical timescales.
We also explore the dependence of the strength of the response on time of merger, merger ratio, and mass of the main halo.
These results provide insight into the physics of halo mergers and their effects on the statistical behaviour of halo properties. 
This study paves the way towards a physical understanding of scaling relations, particularly to how systematics in their scatter are connected to the mass assembly histories of haloes.

\end{abstract}




\section{Introduction}

In the $\Lambda$CDM cosmology \citep[e.g.,][]{planck13,Planck2018_cosmoparams}, structures form hierarchically: objects form on small scales first, and grow through mergers.
Dark matter haloes form around density peaks through gravitational collapse, and provide potential wells for gas to cool and condense, thus forming galaxies \citep[e.g.,][]{whiterees78,blumenthal_etal84}.
The formation and evolution of galaxies, which dominate the observable Universe, is dependent on cosmology through the dark matter field.
Therefore, we must seek to understand the evolution of the dark matter component of the Universe, in order to predict the large-scale structure of the Universe, and interpret cosmological observations.

In the past two decades, with the rapid development of computational capacity, simulations have come to replace analytical models as the primary tool in cosmological studies.
There are two major categories of simulations: dark matter-only, $N$-body simulations \citep[e.g.,][]{boylan-kolchin2009_milleniumII,klypin_etal16,bolplanck2016}, and hydrodynamical simulations that consider baryonic physics \citep[e.g.,][]{schaye2015_eagle,mccarthy2017_bahamas,nelson2018a_TNG}. One of the primary focuses of these simulations is understanding the properties and evolution of dark matter haloes.

A number of properties can be measured for dark matter haloes in simulations, which are physically meaningful in understanding the large-scale structure, modelling the galaxy–halo connection, and exploring the physics of dark matter.
For example, halo concentration is often used as an indicator of the internal structure of haloes \citep[e.g.,][]{nfw97}; it is an important parameter in modelling gravitational lensing \citep[e.g.,][]{umetsu2014}, and provides tests of the particle nature of dark matter \citep[e.g.,][]{bullock2017}.
Halo spin measures the rotation of the halo; it reflects the flow of angular momentum in the gravitational collapse, has systematic alignments with the cosmic web \citep[e.g.,][]{hahn2007b}, and is connected to galaxy spin \citep[e.g.,][]{bett2010} and galaxy size \citep[e.g.,][]{kravtsov13}.
Halo shape measures the triaxiality of haloes; it bears the imprint of the distribution of infalling mass in the halo assembly history \citep[e.g.,][]{vera-ciro2011}, interacts with galaxy formation and evolution \citep[e.g.,][]{debattista2008}, and affects the modelling of weak lensing \citep[e.g.,][]{corless2007}.
Halo position offset reflects the halo's deviation from spherical symmetry, it is often used as relaxation criteria \citep[e.g.,][]{neto07}, and is found to correlate with the past assembly history \citep[e.g.,][]{power2012}.
Halo velocity offset reflects the deviation from dynamical equilibrium, and can be used for similar purposes.
Halo orientation measures the direction of the longest axis in the halo shape, it bridges the orientation of galaxies and the underlying matter field \citep[e.g.,][]{faltenbacher2009}, and provides insight into halo formation in the cosmic web \citep[e.g.,][]{patiri2006}.

The most important halo property, however, is halo mass (or other mass-like properties, such as the depth of the potential well).
Many halo properties and galaxy properties are strongly dependent on halo mass, and their scaling relations with halo mass are instrumental in various cosmological analyses \citep[see, e.g.,][]{moster10,viola2015,wechsler_tinker18}.
It is therefore crucial to understand the systematics in these scaling relations.

The evolution of halo properties with the mass assembly shapes scaling relations, as was discussed extensively in previous works \citep[e.g.,][]{wechsler02,hetznecker2006,wong2012,ludlow2016,chen2020}.
Recently, \citet{mendoza2023} developed a model to predict present-day halo properties from the halo mass assembly history, which is also able to capture correlations between the present-day properties.
As we showed in \citet{Wang2020concentration} (hereafter W20), halo mergers are a potentially essential source of scatter in scaling relations.
In the hierarchical formation, haloes form through mergers, and major mergers are known to trigger violent relaxation in the merging haloes, leading to significant changes of internal structures in the remnant \citep[see, e.g.,][]{mo_vdb_white10}.
In W20, we studied the response of halo concentration to halo mergers in detail.
We found that halo concentration exhibits an intense response to halo mergers, with qualitative universal features and timescales.
We showed that mergers contribute significantly to the scatter in the concentration–mass–formation time relation.

In this work, we expand the previous work and look into the halo anisotropy properties.
We define anisotropy properties as properties that measure the directional variation of the matter and energy distribution in a halo.
These include halo spin, halo shape, halo position offset, halo velocity offset, and halo orientation.
We have discussed some of the applications of these properties above, and will describe them in more detail in the text.
For each anisotropy property, we measure the response in its evolution induced by halo merger activity, interpret the qualitative features in the framework of merger dynamics, explore the factors that affect the responses quantitatively, and discuss the effect of mergers on the co-evolution of properties.

This paper is organised as follows.
In \autoref{sec:sim_cat}, we describe the simulation and catalogues used in the analysis.
In \autoref{sec:methods}, we detail the construction and selection of our samples.
We present and interpret our results in \autoref{sec:results}.
We discuss the implications of our findings in \autoref{sec:discussion} and draw conclusions in \autoref{sec:conclusions}.

\section{Simulation and Catalogues}
\label{sec:sim_cat}

\subsection{IllustrisTNG Simulations}
\label{sec:tng}

The simulation that we use in this work is the TNG300-1 run of the IllustrisTNG simulation suite \citep{marinacci2018_TNG,naiman2018_TNG,nelson2018a_TNG,pillepich2018b_TNG,springel2018_TNG,nelson2019a_TNG}.
IllustrisTNG is a suite of large volume, cosmological, gravo-magnetohydrodynamical simulations run with the \textsc{AREPO} code \citep{springel_2010}.
The simulation suite assumes the Planck 2015 cosmology \citep{planck2016}: $\Omega_{\Lambda,0} = 0.6911$, $\Omega_{m,0} = 0.3089$, $\Omega_{b,0} = 0.0486$, $\sigma_8 = 0.8159$, $n_s = 0.9667$, and $h = 0.6774$.
Of the three simulation volumes, TNG50, TNG100, and TNG300, we use the high-resolution, full-physics run with the largest volume, TNG300-1.
The TNG300-1 run has a box size of $L_{\rm box} = 205\Mpch$, and dark matter and baryon mass resolution of $4\times 10^7\Msunh$ and $7.6\times 10^6\Msunh$ respectively.

In the TNG simulation, haloes are identified with a standard friends-of-friends (FoF) algorithm \citep[e.g.,][]{davis85_fof}, and we adopt the virial masses of haloes provided in the group catalogue.
Subhaloes, which host individual galaxies, are identified with the \textsc{Subfind} algorithm \citep{subfind}.
The merger trees at the subhalo level are constructed with the \textsc{SubLink} algorithm \citep{rodriguez-gomez2015}.
In other words, \textsc{SubLink} assigns descendants and progenitors to the subhaloes identified by \textsc{Subfind}.

\subsection{Terminology of halo catalogue}
\label{sec:terms}

To ensure clarity in the discussion, we explain some key concepts in this subsection, in consistency with IllustrisTNG conventions\footnote{See \citet{subfind,rodriguez-gomez2015} and \url{https://www.tng-project.org/data/docs/specifications/} for more information.}, and highlight the differences from more commonly used definitions.

\subsubsection{Terminology of objects}
\label{sec:terms_obj}

\begin{itemize}

    \item \textit{FoF group}:
    FoF groups are gravitationally bound regions of dark matter overdensities, found with a standard friends-of-friends (FoF) algorithm.
    The boundaries of FoF groups are determined by the particles in the group.

    \item \textit{Halo}:
    In IllustrisTNG, haloes are equivalent to FoF groups.
    However, in our discussion, we use the term to refer to the regions in FoF groups with virial boundaries.
    They inherit the positions of the FoF groups, but have spherical boundaries that enclose a mean density equal to the virial density.

    \item \textit{Subhalo}:
    Subhaloes are self-bound substructures derived within each FoF group with the \textsc{Subfind} algorithm.
    The largest subhalo in the FoF group is the smooth background component of the halo, and the other subhaloes are the smaller substructures in the group.
    This is different from the more commonly adopted definition, where the smooth component is not regarded as a subhalo.

    \item \textit{Central subhalo (\textsc{Subfind})}:
    The central subhalo in the \textsc{Subfind} definition is the subhalo in the FoF group with the highest total number of bound particles/cells, which is usually the most massive subhalo.
    The central subhalo is typically equivalent to the smooth background halo in the common definition.

    \item \textit{Secondary subhalo (\textsc{Subfind})}:
    Secondary subhaloes are subhaloes in the FoF group besides the central subhalo, which are usually referred to as subhaloes in the common definition.

    \item \textit{Central and secondary subhaloes (\textsc{SubLink})}:
    Similar to the \textsc{Subfind} central and secondary subhaloes.
    However, in \textsc{SubLink}, the central subhalo is the subhalo with the most massive history \citep[see Section 2 of][]{de-lucia2007}, which is not necessarily the same as the \textsc{Subfind} central.
    
\end{itemize}

\subsubsection{Terminology of merger trees}
\label{sec:terms_tree}

In this work, we use two types of merger trees, one at the subhalo level, and one at the halo level.
They share some key concepts, which we describe below.

\begin{itemize}

    \item \textit{Merger tree}:
    The merger tree records the links between haloes/subhaloes across different snapshots, and traces the history of how smaller objects merge to form larger objects.

    \item \textit{Descendant}:
    A halo/subhalo evolves or merges into its descendant halo/subhalo, each halo/subhalo can have no more than one immediate descendant in the following snapshot.
    The descendant of a subhalo is chosen from candidates that have common particles with it, with a merit function based on the binding energy of common particles \citep[see Section 3.1 of][for details]{rodriguez-gomez2015}.
    The descendant of a halo inherits its central subhalo.

    \item \textit{Progenitor}:
    Progenitor haloes/subhaloes evolve or merge into the descendant halo/subhalo, a halo/subhalo can have any number of progenitors.
    Halo/subhalo $A$ is a progenitor of halo/subhalo $B$ if and only if $B$ is the descendant of $A$.

    \item \textit{Main and secondary progenitors}:
    The main progenitor subhalo is the progenitor with the most massive history \citep{de-lucia2007}.
    In general, the main progenitor halo is the host of the main progenitor subhalo, and we discuss this in more detail in \autoref{sec:tree_construct}.
    The other progenitors are considered secondary progenitors.
    
    \item \textit{Main branch}:
    The main branch of the tree consists of the main progenitors along the history of the halo/subhalo.
    A mass assembly history has secondary branches besides the main branch, which also contribute to the growth of the halo.
    
\end{itemize}

\subsection{Halo properties of interest}
\label{sec:props}

In this work, we investigate properties of the \textit{dark matter component} of the haloes beyond what is available in the IllustrisTNG \textsc{SubLink} catalogues. 
For each halo, we take all particles of its FoF group and select the subset of particles bound to the halo. Unbinding is performed by computing kinetic energies relative to the FoF centre-of-mass velocity of the halo and potential energies are calculated by assuming an NFW profile \citep{navarro_1997}, fitting for the concentration relative to the position of the most-bound particle, and computing escape velocities.
We then use this bound subset to calculate halo properties.
All procedures, including that of unbinding and calculating the properties mentioned below, will be described in detail in an upcoming work \citepinprep{Anbajagane}.

We focus on the anisotropy properties of haloes.
These properties measure the level of directional variation of a halo's density and energy distribution, and provide insight into its dynamical state.
We reproduce the key pieces of the calculation for the properties of interest below:

\begin{itemize}
    \item $M_{\rm vir}$, {\em the virial mass} - $M_{\rm vir}$ is defined as the mass within the virial radius, $R_{\rm vir},$ and the virial radius is the radius at which the halo's enclosed density first becomes some target value, $\rho_{\rm ref}$:
    \begin{equation}
        M_{\rm vir} = M(<R_{\rm vir}) = \rho_{\rm ref} \frac{4\pi}{3}R_{\rm vir}^3.
    \end{equation}
    We use the definition of $\rho_{\rm ref}$ computed in \citet{bryan_norman98}. We also define the virial velocity as the circular velocity of a test particle at the virial radius:
    \begin{equation}
        V_{\rm vir} = \sqrt{GM_{\rm vir}/R_{\rm vir}}.
    \end{equation}
    \item \textit{$\spinB$, the Bullock spin} \citep{bullock01} - 
    $\spinB$ is defined as 
    \begin{equation}
        \spinB=\frac{|\vec{J}|}{\sqrt{2}\Mvir\Rvir\Vvir}
    \end{equation}
    where $\Mvir$ is the virial mass, $\Rvir$ is the virial radius, $\Vvir$ is the virial circular velocity, and
    \begin{equation}
        \vec{J} = \sum_{i = 1}^N m_i (\vec{x}_i - \vec{x}_{\rm cen}) \times (\vec{v}_i - \vec{v}_{\rm core}).
    \end{equation}
    Here $\vec{x}_{\rm cen}$ is the halo centre, defined as the most bound particle (of any type) in the halo, and $\vec{v}_{\rm core}$ is the core velocity, defined as the mean velocity of all particles (of the type of interest, i.e. DM) within $r < 0.1\Rvir$. In case there are fewer than 100 particles within $0.1\Rvir$, we use the 100 particles nearest to the centre of the halo, to compute $\vec{v}_{\rm core}$.
    
    \item \textit{$c/a$, the axis ratio} -
    $c/a$ is the ratio between the shortest and longest axis of the halo shape, sometimes denoted as $s$. This is estimated by computing the reduced mass-inertia tensor \citep{zemp2011},
    \begin{equation}
    \mathcal{M}_{ij} = \sum_{k = 0}^{N} \frac{x_{k, i} \,x_{k, j}}{r^2_{{\rm ell}, k}}
    \end{equation}
    where $r^2_{{\rm ell}, k}$ is the elliptical radius to the $k$th particle,
    \begin{equation}
    r^2_{\rm ell} = \tilde{x}^2 + \frac{\tilde{y}^2}{b^2} + \frac{\tilde{z}^2}{c^2}
    \end{equation}
    with $\tilde{x}$, $\tilde{y}$, $\tilde{z}$ as the coordinates along the eigenvectors of the halo, and $b$ and $c$ are the intermediate and minor axis lengths. This choice of setting $a = 1$, is equivalent to setting $a = \Rvir$. So the semi-major axis is always given by $\Rvir$. The axis lengths $a,\,b,$ and $c$ are the square roots of the eigenvalues of this tensor. 
    We perform this calculation iteratively. We first make an estimate of the shapes using all particles within $\Rvir$ (i.e. a spherical volume cut). Next, we iteratively redo our particle selection as $r^2_{\rm ell} < \Rvir^2$, with $r^2_{\rm ell}$ measured using the shapes and eigenvectors computed in the previous iteration of the measurement. Once the inferred $s=c/a$ and $q=b/a$ estimates both vary by less than $<1\%$ from its previous estimate, we say the calculation is converged.
    
    \item \textit{$\xoff$, the position offset} -
    $\xoff$ is the distance between the location of the most gravitationally bound particle and the centre of mass.
    We normalise $\xoff$ by the virial radius $\Rvir$, and consider the dimensionless property $\xoffnorm$.
    
    \item \textit{$\voff$, the velocity offset.} -
    $\voff$ is the difference between the core velocity, $\vec{v}_{\rm core}$, and the mean velocity.
    We normalise $\voff$ by the virial circular velocity, $\Vvir$, and consider the dimensionless property $\voffnorm$.
    
\end{itemize}

All of these properties besides mass reflect the state of anisotropy of haloes, and are closely related physically.
In addition to these properties, we also include halo orientation.

\begin{itemize}
    \item \textit{$\cos\theta$, the change in halo orientation} -
    We measure the halo orientation through the direction of the major axis. This direction is simply the eigenvector corresponding to the eigenvalue $a$ mentioned above, and is obtained by diagonalising the reduced inertia tensor.
    Specifically, we choose a time of reference, and set the major axis at this time as as the orientation of reference.
    We then measure the angle $\theta$ between the major axis at each snapshot and the reference orientation, and record the value of $\cos\theta$.
\end{itemize}

\section{Methods}
\label{sec:methods}

To study the behaviour of dark matter haloes in response to mergers, we need to define a physically meaningful sample of merger events.
In this section, we describe the compilation of our data sample.

\subsection{Base halo sample selection}
\label{sec:halo_samp}

\begin{figure}
    \centering
    \includegraphics[scale=0.49, trim=0 0.9cm 0 0, clip]{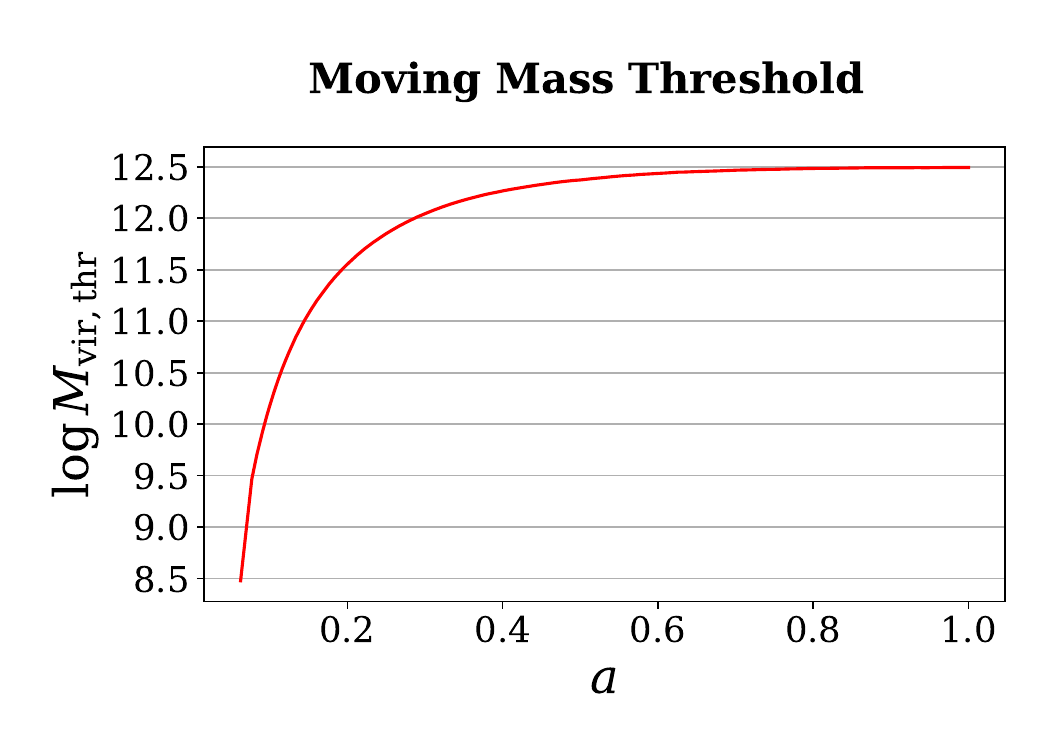}
    \caption{The mass threshold used to select our base halo sample, as a function of scale factor.
    This mass threshold corresponds to the 10000 haloes with the highest virial masses in each snapshot.}
    \label{fig:mthr}
\end{figure}

We first define a sample of haloes in TNG300-1, in whose assembly histories the mergers occur.
At each snapshot, we select a fixed number of haloes with the highest virial masses, such that the halo sample is mass-complete with a moving mass threshold as a function of cosmic time.
We select the 10000 most massive haloes at each snapshot, which results in a mass threshold of approximately $10^{12.5}\Msunh$ at $a=1$, and $10^{12}\Msunh$ at $a=0.3$.
We show the mass threshold as a function of scale factor in \autoref{fig:mthr}.
This is a practical choice considering the volume of the simulation, as the number of haloes increase rapidly when the mass threshold is lowered.
We discuss the implications of this moving mass threshold treatment in \autoref{sec:mthr_bias}.
We note that the sample is not exactly consistent throughout cosmic time, in other words, the haloes selected at any two different snapshots are not guaranteed to be the same.

\subsection{Halo assembly history construction}
\label{sec:tree_construct}

We then construct the mass assembly histories of the haloes that we have selected.
This is a nontrivial procedure, because the \textsc{SubLink} merger tree provided by TNG is based on subhaloes instead of haloes.
In the subhalo-based merger trees, a merger happens when two or more subhaloes share the same descendant, and are no longer identified as independent objects.
This typically happens between subhaloes in the same halo, and differs from the halo merger history that we are interested in.
We convert the links between subhaloes to links between haloes with the following method.
We develop a new Python package, \textsc{TreeHacker}, to perform the following navigation and manipulation of the \textsc{SubLink} catalogues.
We describe the package in more detail in \citetinprep{Wang}.

\subsubsection{Main branch history construction}
\label{sec:mb_tree}

For each given halo, we locate its central subhalo as identified by \textsc{SubLink}.
We extract the main branch progenitors and descendants of the central subhalo from \textsc{SubLink}, and locate the haloes that host these subhaloes.
We treat these haloes as the progenitor and descendant haloes of the given halo, which will form the preliminary main branch history.

We further process the halo histories based on their mass evolution, in order to exclude any splashback haloes \citep[e.g.,][]{diemer_kravtsov14}, i.e., subhaloes which have temporarily orbited outside the FoF boundaries of their hosts. Most subhaloes become splashback haloes close to the time of their first apocentre.
Although they will appear as primary haloes in the underlying FoF catalogue, splashback haloes are subhaloes, and will experience mass loss and profile truncation due to their more massive host in the same way that all subhaloes do. This makes them a distinct physical category from host haloes and means that their internal properties are systematically different from ordinary host haloes, too. These facts make it important to remove them from our sample.

Ideally, we would identify splashback haloes by searching through the the main branches of central subhaloes and identifying times when they were secondary subhaloes. However, we find that the primary/secondary classification within a group experiencing a major merger often switches back-and-forth, causing this procedure to over-classify distinct host haloes as splashback haloes, as has been previously reported for other subhalo finders and merger tree codes \citep{mansfield_kravtsov_2020}. There are some existing methods for accounting for these types of errors \citep{mansfield_2023}, but in this paper, we opt for an approach based on mass evolution.

The mass accretion histories of splashback haloes are marked by sudden and significant pairs of increases and decreases in their group mass, sometimes by several orders of magnitude.
We identify these signatures, and truncate the history of a splashback halo at its infall into the more massive host.
We outline the procedure as follows.
First, we locate drops in the subhalo's group mass which are larger than 0.5 dex between adjacent snapshots, as an indication of the subhalo escaping from a more massive host; we remove the snapshots after the escape.
Second, we identify the last snapshot in the remaining history where the main branch subhalo is the central subhalo of the host halo, and discard all later snapshots; this excludes the history after the halo falls into the more massive host.

There may be overlaps in the different halo histories obtained at this point.
For example, when two haloes with similar masses merge, the descendant halo is included in both histories.
We compare the histories that have overlapping portions, and identify the one that survives for the longest time afterwards.
We choose to keep the longest surviving tree unchanged, and truncate the other histories at the earliest point of overlap, so that the time span of the main branch history is maximised.
In other words, we choose the progenitor halo whose central subhalo survives for the longest time (which, we find, is typically the more massive progenitor), as the main branch progenitor for the descendant.

With the above described procedure, we construct a set of main branch histories of the mass-complete halo sample.
We will locate the mergers that happen to these haloes along their main branches, measure the halo properties during the relevant time span, and track their evolution in response to the mergers.

\subsubsection{Merging halo identification}
\label{sec:merging_tree}

For a given main branch history, we now identify the haloes that merge into it.
At each snapshot along the main branch history, we select a maximum of 20 most massive subhaloes in the main halo, and retrieve their main progenitors from \textsc{SubLink}.
If a progenitor subhalo is part of a host halo other than the main branch progenitor halo that is within the sample defined in Section \ref{sec:halo_samp}, this other host halo is recorded as a candidate merging halo.

One rare edge case can occur if a splashback halo orbits outside of its original host and is immediately accreted onto a second host in the next snapshot. This would could cause the splashback halo's original host to appear to be a progenitor of the splashback halo's second host.
We deal with this issue by imposing the additional condition that the subhalo's mass must be above a significant fraction of its original host's group mass at the snapshot before it joins the FoF group, for its host to be considered a merging halo.
The threshold fraction we adopt is 0.5 dex.
Above this threshold, the splashback scenario is highly unlikely.
Furthermore, the merging halo must enter the virial boundary of the main halo at some point during its history to be included in the sample, instead of just being a member of the same FOF group.
In our following analyses, we also apply thresholds to the mass ratio between the main and merging haloes.

The product of this process is a set of haloes that merge with the main branch haloes from the mass-complete sample, from which we select merger events that we will eventually stack and analyse.

\subsection{Dynamical time}
\label{sec:tdyn}

Halo mergers are dynamical processes, and the local dynamical timescale is a relevant timescale in studying mergers.
We introduce this measure of time before discussing our selection of mergers.
The dynamical time is the time required to orbit across an equilibrium dynamical system, in our case a halo.
We adopt the definition of the dynamical time in \citet{mo_vdb_white10},
\begin{equation}
    \tdyn(t)=\sqrt{(3\pi)/(16G\bar{\rho}(t))},
    \label{eq:tdyn}
\end{equation}
where $G$ is the gravitational constant and $\bar{\rho}(t)$ 
is the mean density of the system, which is the virial density of haloes.
With a given cosmology, the dynamical time $\tdyn(t)$ is dependent on the 
cosmic time through $\bar{\rho}(t)$.

Following our definition in W20 \citep[see also][]{Jiang_vdB16}, we use the quantity $T_{\rm dyn}$ to measure the time between two epochs in units of the dynamical time, as
\begin{equation}
    T_{\rm dyn}(t(a);t(a_{\rm ref})) = \int_{t(a_{\rm ref})}^{t(a)}\frac{dt}{\tdyn(t)},
    \label{eq:Ntau}
\end{equation}
where $a_{\rm ref}$ is the epoch of reference, $a$ is the epoch of interest, and $t(a_{\rm ref})$ and $t(a)$ the corresponding cosmic times. 
A quantity of particular interest is the number of dynamical times with respect to a given merger, $\Tmerger$.

\subsection{Merger sample selection}
\label{sec:merger_selection}

We use several different samples of mergers for different purposes in our analyses, selected with different criteria in three dimensions, each of which we describe below.

\subsubsection{Time of merger, $\amerger$}
\label{sec:crit_amerger}

We define the time of merger as the scale factor at which the centre of the secondary halo crosses the virial boundary of the main halo.
In this work, we consider mergers that have intermediate values of $\amerger$.
We exclude early mergers because haloes at early times are typically not well resolved, and we also exclude late mergers to allow for sufficient time intervals between the mergers and the present-day snapshot.
We divide $\amerger$ into three bins: 
\begin{align*}
    0.28 \leq\amerger < 0.37;\\ 
    0.37 \leq\amerger < 0.50;\\ 
    0.50 \leq\amerger < 0.67.
\end{align*}
Each of these bins spans approximately 1.5$\tdyn$, and the latest time of merger included in the last bin is approximately 2$\tdyn$ from the present-day snapshot.

We impose a further requirement on the mergers to include in the analysis, that the main halo is an independent host throughout the time span of $-1\leq\Tmerger\leq2$ around $\amerger$.
This is to ensure sufficient tracking of the evolution of haloes during and after mergers. This selection will remove some haloes which are in the infall regions of larger haloes. This biases our sample towards faster-accreting haloes, although it is likely a sub-dominant effect compared to the accretion history constraints imposed by our mass threshold.

\subsubsection{Merger ratio, $M_2/M_1$}
\label{sec:crit_rmerger}

One of the defining features of a halo merger is the mass ratio between the two merging progenitors.
We measure the ratio between the virial masses of the two merging haloes, at the snapshot immediately before the secondary halo joins the FoF group of the main halo.
At this time, the virial masses of both haloes are relatively well-defined.
The main progenitor is more massive than the secondary progenitor by definition, in other words, $M_2/M_1<1$.
We divide $M_2/M_1$ into three logarithmic bins: 
\begin{align*}
    0.032\leq M_2/M_1<0.10;\\
    0.10\leq M_2/M_1<0.32;\\
    0.32\leq M_2/M_1<1.00,
\end{align*}
which correspond to 0.5 dex intervals.
The highest bin has a threshold ratio close to $1:3$, a ratio often adopted to define major mergers.

\subsubsection{Previous halo mass, $M_{-1}$}
\label{sec:sec_prevmass}

The final feature we use to characterise mergers is the virial mass of the main halo at one dynamical time before the time of merger, i.e., at $\Tmerger=-1$.
We denote this previous mass as $M_{-1}$.
At this time the secondary halo has typically not started interacting with the main halo, and it is reasonable to compare between two halo populations with similar previous masses that differ in the ensuing merger history.
We consider logarithmic bins of 0.5 dex of $M_{-1}$.
In the main text, we study mergers with
\begin{align*}
    10^{12}\Msunh\leq M_{-1}<10^{12.5}\Msunh;\\
    10^{12.5}\Msunh\leq M_{-1}<10^{13}\Msunh.
\end{align*}
In \autoref{sec:mthr_bias}, we examine bins in the lower mass range, $10^{11}\Msunh\leq M_{-1}<10^{12.5}\Msunh$.

\subsection{Random sample selection}
\label{sec:random_selection}

To understand the baseline evolution of halo properties, against which the merger-induced evolution is to be compared, we select random segments of halo evolution from the main branch histories constructed in \autoref{sec:mb_tree}.
We treat these segments in a similar way to the merger sample: we assign a time of reference, $a_{\rm ref}$, in analogy to $\amerger$ for the merger sample, and require that the halo is present as an independent host throughout the time range of $-1\leq T_{\rm dyn;ref}\leq2$.
As with mergers, we also characterise the random segments with the previous mass $M_{-1}$, measured at $T_{\rm dyn;ref}=-1$.
On the other hand, the random segments have no merger ratios associated with them.

We define a corresponding random sample for each merger sample.
Specifically, the range of $\amerger$ matches the range of $a_{\rm ref}$, and the two samples share the same range of $M_{-1}$.
In the following analysis, we will stack the property evolution for the merger sample and the corresponding random sample respectively, and examine the differences induced by merger events through comparison.

However, the differences observed between the merger and random samples cannot be completely attributed to single mergers. The presence of mergers typically implies higher accretion rates, and halo accretion raters are correlated with environment \citep[e.g.][]{gao_etal05}, meaning that the merger sample and and random sample are taken from different environments. This means that, for example, at low masses, haloes in our random sample would be experiencing more tidal truncation from their large-scale environment than our merger sample \citep{mansfield_kravtsov_2020}. There is no trivial way to account for such systematic biases. We expect the single mergers to be the predominant source of the signal post-merger, but such differences may cause the two samples to differ during pre-merger times and may limit the interpretability of, say, the ratio between a quantity measured for the random sample and the same quantity measured in the merger sample.

\section{Results}
\label{sec:results}

\subsection{Basic dynamics of mergers}
\label{sec:dynamics}

We start by describing the basic dynamics of a merger between two haloes.
In reality, the physics of halo mergers is highly complicated, with collisionless mechanisms such as tidal stripping and dynamical friction in act, as well as the baryonic interactions that accompany the following galaxy merger.
However, we focus on the orbital behaviour of a merger, which, as we will show later, is sufficient to explain our main qualitative findings.
In general, when two haloes merge, the secondary, less massive halo falls into the more massive main halo, loses mass and energy while orbiting within it, and eventually dissolves.

\begin{figure}
    \centering
    \includegraphics[scale=0.56, trim=1cm 1.5cm 1cm 1.5cm, clip]{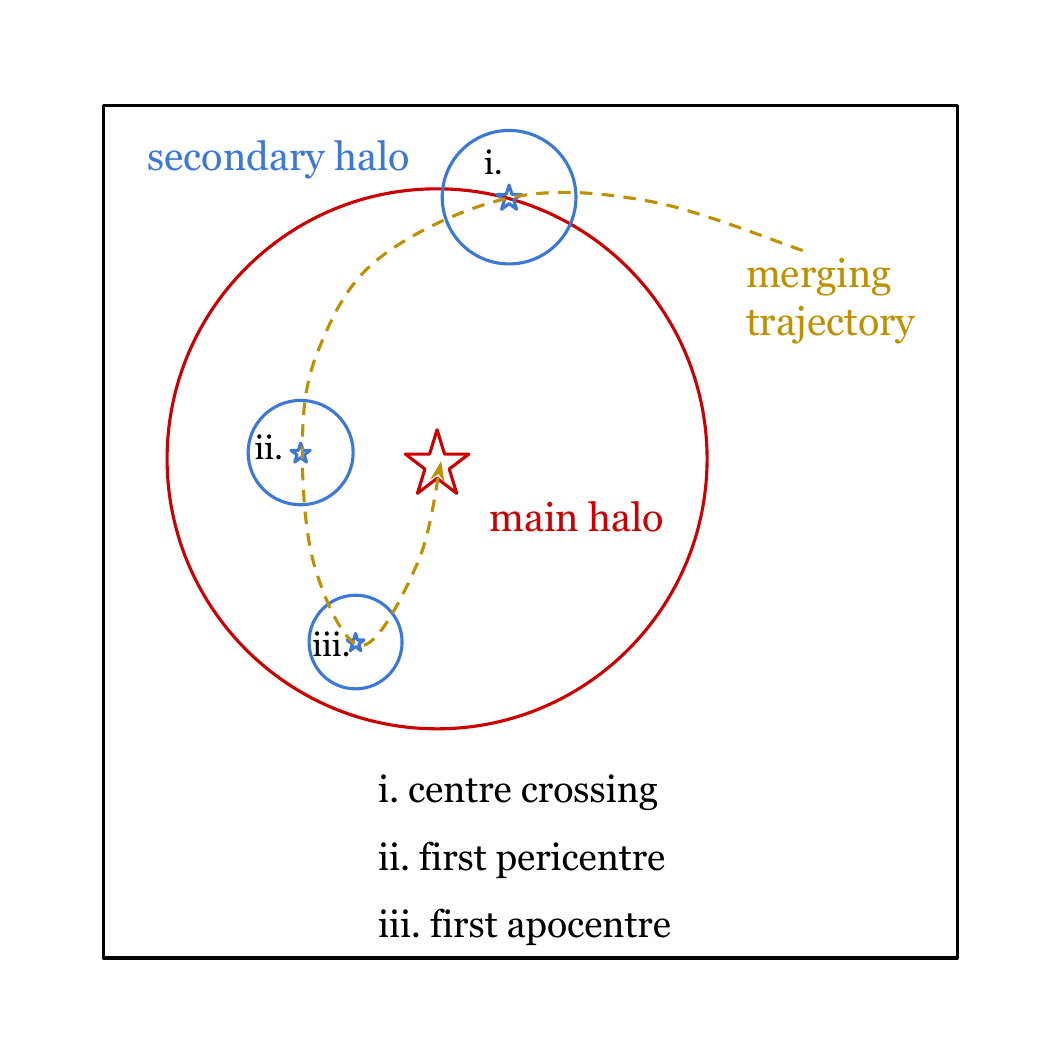}
    \caption{Characteristic stages of the merger.
    We schematically illustrate the orbit of the secondary halo with respect to the main halo during and after the merger.
    We label the three most important dynamical stages in the process, which are connected to the merger responses of halo properties.
    Note that the increase in the radius of the main halo is not reflected in the illustration.}
    \label{fig:cartoon}
\end{figure}

We show a simplified version of the merger process in \autoref{fig:cartoon}.
We use several characteristic stages in the orbit to outline the dynamical process during the merger.
We approximately treat haloes as objects with spherical boundaries, and our discussion is based on the virial radii of haloes.
These approximations do not represent the physical reality during a merger, and we adopt them for simplicity.

\begin{itemize}

    \item {\it Centre crossing}: 
    The time when the centre of the secondary halo crosses the virial radius of the main halo.
    This is our definition of the time of merger, as described in \autoref{sec:merger_selection}.

    \item {\it First pericentre}:
    The point of closest approach in the initial infall.
    The distance between the centres of the secondary and main haloes monotonically decreases before the first pericentre, and starts to increases after it.

    \item {\it First apocentre}:
    The farthest point from the centre of the main halo reached by the secondary halo, after passing the first pericentre.
    At this point, the secondary halo turns back and starts to approach the centre of the main halo again.
    
\end{itemize}

We note that after the centre crossing, the secondary halo has merged into the main halo, and is no longer an independent halo. In some places, however, we continue to refer to descendant subhalo as the secondary halo for convenience.
The transfer of mass and energy is more rapid and more violent for major mergers, where the two haloes have similar masses.
In major mergers, the secondary halo is typically disrupted rapidly over a small number of orbits, whereas less massive secondary haloes may continue to orbit for an extended period of time within the main halo \citep[see, e.g.,][]{mo_vdb_white10}.

\subsection{Orbits in dynamical times}
\label{sec:orb_tdyn}

For a quantitative study of the orbital behaviour of the secondary halo, we examine the timescales associated with the characteristic stages in the orbit.
We measure the times of the first pericentre and first apocentre with respect to the centre crossing.
We select a merger sample with
\begin{align*}
    &0.5\leq\amerger<0.67,\\
    &0.32\leq M_2/M_1<1.00,\\
    &12\leq\log M_{-1}<12.5,
\end{align*}
and make measurements on the stacked orbit of the secondary haloes in these mergers.
We choose the latest bin of $\amerger$, which maximises the temporal resolution in units of dynamical times.
We choose the bin of most major mergers with the highest $M_2/M_1$.
We also choose the mass range containing the most mergers that satisfy the first two criteria, to improve the statistics.
We examine other samples of mergers as well, and find that the stacked orbit has a clear dependence on the merger ratio, which we explore in more detail in \autoref{sec:diff_orbit_time}.

\begin{figure}
    \centering
    \includegraphics[scale=0.57, trim=0 0.5cm 0 0, clip]{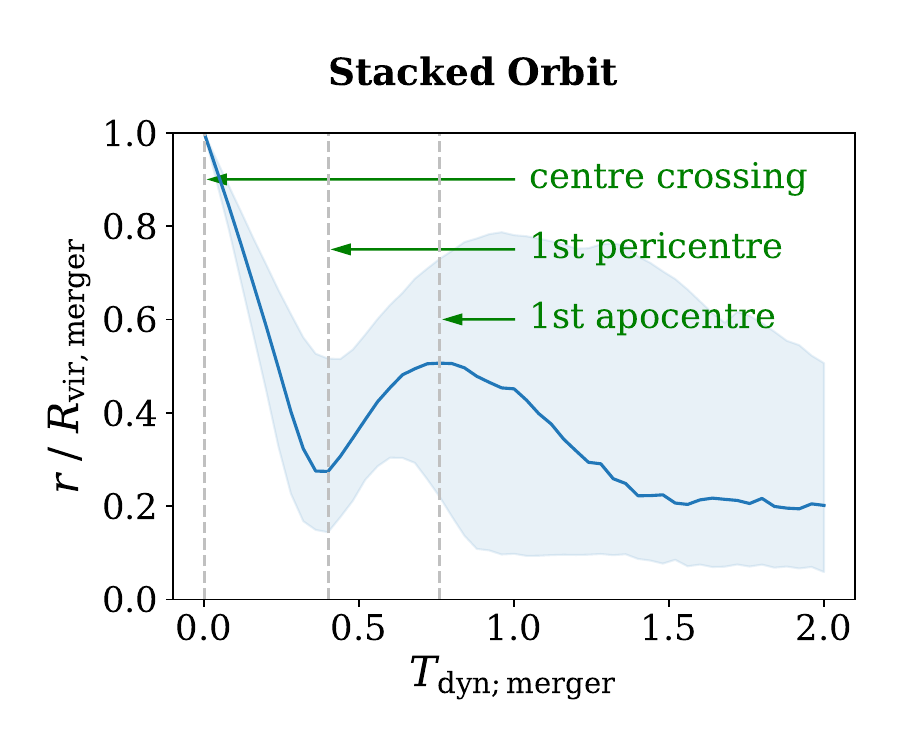}
    \caption{Stacked orbit radius of the descendant subhalo of the secondary halo with respect to the main halo, for a sample of major mergers.
    The orbit radius is normalised by the host halo virial radius, and plotted as a function of the dynamical time since the time of merger.
    We show the median orbit as well as the 16-84th percentile range.
    We measure and mark the timescales of the three characteristic stages in the merger process with vertical dashed lines.
    The orbit reflects the typical trajectory of the secondary halo in mergers.}
    \label{fig:orbit_radius}
\end{figure}

In \autoref{fig:orbit_radius}, we show the median evolution of the distance between the centres of the two haloes, i.e., the radius of the secondary halo's orbit, normalised by the virial radius of the main halo at the time of merger.
The evolution of the radius in consistent with the scenario described in \autoref{sec:dynamics}.
Upon crossing the virial boundary of the main halo at $\Tmerger=0$, the radius decreases, reaches a local minimum, which is the first pericentre, temporarily increases and decreases again, peaking at a local maximum, the first apocentre.
The stacked orbit for this major merger sample does not show signals of other pericentres and apocentres.
The location of the first pericentre (local minimum) and the first apocentre (local maximum) are marked by vertical dashed lines and labelled in the figure. 
We find that in the stacked orbit, the first pericentre occurs at $\Tmerger=0.40$, and the first apocentre occurs at $\Tmerger=0.76$, marked by vertical dashed lines in the figure.

We also show the range that corresponds to the 16-84th percentiles as a shaded region, to provide an estimate of the scatter in the sample distribution.
While the upper and lower bounds do not correspond to specific orbits, the shape of the 16-84th percentile range suggests that individual orbits exhibit varying timescales.
The time of the first pericentre is relatively universal, whereas the time of the first apocentre spans a broader range, and shows a correlation between smaller apocentre radii and earlier apocentre times.
While all haloes in the same epoch share the same overall dynamical time, haloes have higher densities in their inner regions, leading to shorter local dynamical times at smaller radii.
The secondary haloes that have smaller orbit radii therefore reach the apocentre sooner than the secondary haloes that have larger orbit radii.
The different orbits may be due to different levels of dynamical friction experienced by the secondary haloes.

We now discuss the caveats that arise from our stacking procedure.
For each merger, we first linearly interpolate between available snapshots, with intervals of $\Delta T_{\rm dyn}=0.04$, because mergers happen at different times, and the orbits are sampled at different $\Tmerger$ points.
Our results are therefore limited by the temporal resolution of the simulation.
At each interpolated $\Tmerger$, we then take the median and percentiles for all the secondary haloes that are still present as subhaloes, because secondary haloes are completely disrupted at different times.
This approach inevitably leads to a survivor bias that favours orbits that last longer in time, which are likely also the orbits that have larger radii and longer timescales.
As a result, the measured pericentre and apocentre are likely biased towards later times, and the apocentre, which occurs later, is subject to a stronger bias.  Similar survivor bias effects have been noted in other studies of subhalo disruption \citep{han_2016,diemer_2023}. There exist techniques for correcting for this effect for properties which change monotonically with time \citep{mansfield_2023}, but we are unaware of any method which can correct for this effect for non-monotonic quantities.

\subsection{Halo property response to mergers}
\label{sec:response}

\begin{figure*}
    \centering
    \includegraphics[scale=0.55]{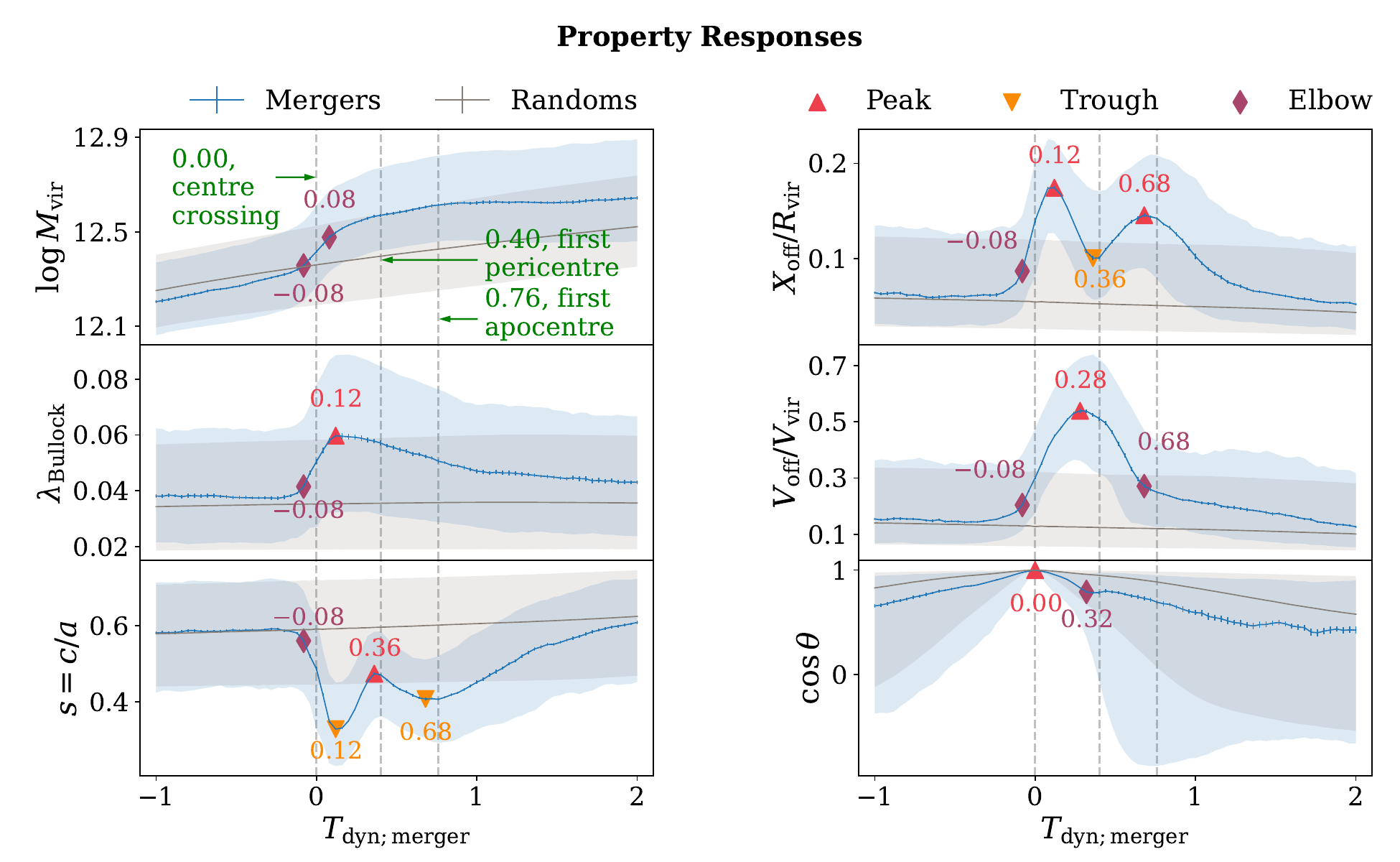}
    \caption{Stacked response of each property for the same sample of major mergers as used in \autoref{fig:orbit_radius}.
    Blue is the merger sample, and gray the random control sample.
    The banded region illustrates the scatter.
    There are error bars corresponding to the bootstrapped error on the median, which are roughly the size of the line.
    The merger sample behaviour has a statistically significant deviation from the random sample. 
    Symbols indicating local maxima (up triangles), local minima (down triangles), and ``elbows'' (extrema in the second derivative; diamonds) are also shown to help guide the eye.
    Discussion of the curves shown in each panel can be found in Sections \ref{sec:mvir_response} - \ref{sec:orientation_response}.}
    \label{fig:uncertainty}
\end{figure*}

In this subsection, we present the stacked merger responses of the halo properties listed in \autoref{sec:props}.
We use the same merger sample as in \autoref{sec:orb_tdyn}, and select the corresponding random sample to have the same range of time and previous mass\footnote{We test that the results from different merger samples are qualitatively similar, thereby confirming the universality of property responses that we found in W20 (see also \autoref{fig:response_dependences}) for a presentation of universality in the halo properties presented in this work.}.

In \autoref{fig:uncertainty}, we show the evolution of each halo property for both the stacked merger sample and the stacked random sample. The solid lines represent the median evolution of each sample. 
Similar to \autoref{sec:orb_tdyn}, for each merger, we linearly interpolate the properties between available snapshots, with intervals of $\Delta T_{\rm dyn}=0.04$.
We illustrate the 16-84th percentile range with shaded regions and the bootstrap standard error of the median with error bars, as labelled above the top left panel.
The scatter is relatively large and comparable to the amplitudes of the responses, whereas the error bars are very small and barely visible outside of the line width.
We mark the times of the centre crossing, first pericentre, and the first apocentre with vertical dashed lines.  We measure the latter two quantities as described in \autoref{sec:orb_tdyn}.
In the figure, we mark the locations of peaks (red triangle) and troughs (upside down orange triangle) in the property responses.
We also mark the most prominent elbow-like (purple diamond) features ---  sharp turns --- indicated by local extrema in the second derivatives of the response functions.
The peaks, troughs, and elbows correspond to characteristic features in the responses.
The different marks are labelled above the top right panel.

We find that for each property, the evolution of the merger sample is significantly different from the random sample.  Specifically, the median property response and their error bars, roughly the size of the line widths, have clear qualitative differences between each sample. The deviations demonstrate the material impact of mergers on halo properties.
The random sample evolution in grey reflects the secular evolution of the general halo population over considerable portions of cosmic time. The comparison with the blue curve shows that the merger-induced response is significant in amplitude over the course of 1-2 dynamical times. Mergers, therefore, lead to significant deviations of these halo properties over these timescales.

We can associate the characteristic features in each property response with the dynamical processes in a merger.
Below, we discuss the features in the qualitative response for each property in relation to the characteristic stages of mergers.
We report estimates of the timescales associated with the peaks, troughs, and elbows corresponding to characteristic response features.
However, we caution against a purely quantitative interpretation of these results for the same reasons why we do not provide functional fits. The accuracy of our measurements is limited by the temporal resolution of both the simulation and the interpolation procedure. And, the details of the quantitative behaviour differ for different merger samples (e.g. see \autoref{fig:response_dependences}).

\subsubsection{Mass response}
\label{sec:mvir_response}
The merger response of $\Mvir$ is as expected.
The median $\Mvir$ generally increases over time for both the random sample and the merger sample.
Compared to the random sample, the merger sample shows a period of rapid growth around the time of centre crossing, which corresponds to the accretion of the secondary halo onto the main halo.
We find elbow features in the merger response at $\Tmerger=-0.08$ and $0.08$, which can be associated with the start and end of the rapid growth period.
While we cannot unambiguously define the time span of the accretion, the majority of the particles from the secondary halo transfer to the main halo during this period.

We note that although the random sample uses the same $M_{-1}$ mass bin as the merger sample, it does not have the exact same distribution of $M_{-1}$ values within that bin.  The differing distributions lead to a small offset in the stacked $\Mvir$ of each sample before the merger begins.
The pre-merger evolution of the other properties also reflect this slight offset.
However, we do not expect this to impact our qualitative conclusions.

\subsubsection{Spin response}
\label{sec:spin_response}
The $\spinB$ evolution of the random sample is almost flat. On the other hand,  the merger sample shows a clear response.
The merger sample exhibits a dramatic increase in $\spinB$ that starts at $\Tmerger=-0.08$, same as the start of the rapid mass growth, and reaches a peak at $\Tmerger=0.12$.
The spin then decreases, with the rate of decrease slowing as the value of $\spinB$ approaches the baseline indicated by the random sample.

Note, the fast increase in $\spinB$ happens around the time of merger.  This is due to the large amount of angular momentum that the secondary halo brings in with respect to the main halo \citep{vitvitska_2002}.
The uptick of $\spinB$ approximately coincides with the period of rapid mass growth seen i/n the $\Mvir$ evolution, lasting slightly longer, as the mass inflow continues.
We attribute the initial decrease that follows to a combination of three effects.
First, during mergers, a significant fraction of the host halo's bound mass flows outside the virial radius \citep{kazanrzidis_2006} while still remaining on bound orbits. This outflowing mass is preferentially particles with high energy and high angular momentum, which can lead to a decrease in the total angular momentum of the halo.
Second, we measure the angular momentum with respect to the halo's most bound particle. Throughout the merger the halo's centre of mass becomes meaningfully offset from the most bound particle (see Section \ref{sec:xoff_response}) and the centre of mass can have a non-zero tangential velocity relative to the most bound particle. This means that in addition to the usual spin angular momentum, $\spinB$ is also temporarily measuring the orbital angular momentum of the halo's centre of mass relative to the most bound particle.
Finally, while the background smooth accretion of mass onto the main halo brings in additional mass, the added amount of net angular momentum is disproportionately small compared to the merger; the dynamics of relative smooth accretion dilutes the spin \citep{vitvitska_2002}.
These effects become less important as spin continues to decrease and the secondary halo disrupts, reducing the rate of decrease in $\spinB$.

\subsubsection{Shape response}
\label{sec:shape_response}

The two axis ratios, $s=c/a$ and $q=b/a$, behave similarly in the qualitative sense, and our discussion of $c/a$ applies to $b/a$ as well.
Lower values of axis ratios correspond to less spherical shapes.
The random sample experiences a slow increase in $c/a$ over time, as haloes tend to become more spherical as accretion rates decrease with decreasing redshift.
In comparison, for the merger sample, the mass brought in by the secondary halo causes a temporary elongation in the shape of the main halo.
It is natural to expect that the elongation in shape is stronger when the two density peaks are farther apart, in other words, when the secondary halo is farther away from the centre of the main halo.

The axis ratio starts to rapidly decrease at $\Tmerger=-0.08$, and reaches its minimum at $\Tmerger=0.12$.
This coincides with the phase of rapid increase in $\spinB$, as the elongation of the halo and the inflow of angular momentum happen simultaneously.
$c/a$ then increases as the secondary halo moves toward the first pericentre, and reaches a local maximum at $\Tmerger=0.36$, near the pericentre.
This local maximum is still lower than the initial value, showing that the mass from the secondary halo elongates the main halo even at the pericentre, as expected from a non-zero impact factor.
Another period of decrease follows as the secondary halo moves toward the first apocentre, and reaches a local minimum at $\Tmerger=0.68$, near the apocentre. The axis ratio during this second minima is larger than the first minima, likely due to some combination of mass loss in the secondary and dynamical friction leading to a small apocentre.
Finally, the median response undergoes another stage of increase, as the secondary halo turns back from the first apocentre and moves toward the centre, before it is completely disrupted.

We note that the local maximum occurs slightly earlier than the pericentre, and the second local minimum occurs slightly earlier than the apocentre.
A possible explanation is the survivor bias discussed in \autoref{sec:orb_tdyn}, which causes the time of the pericentre and apocentre to be overestimated.
A pair of competing effects may also contribute to the offset of the second local minimum: increasing elongation as the secondary halo moves away from the centre, and suppressed elongation as the mass bound to the secondary halo decreases.

\subsubsection{Position offset response}
\label{sec:xoff_response}

The offset between the centre of mass and the most bound particle, normalised by the virial radius, $\xoffnorm$, slowly decreases for the random sample with the secular relaxation process.
For the merger sample, $\xoffnorm$ shows features very similar to $c/a$, though in the opposite sense.
This can be intuitively understood with the orbit of the secondary halo: when the secondary halo moves closer to the centre of the main halo, the position offset decreases, and the halo becomes more spherical.
The two properties are therefore strongly correlated physically.
There are two local maxima and one local minimum in the $\xoffnorm$ response, corresponding to the two local minima and one local maximum in $c/a$, with similar timescales, as marked in the figure.
These findings are in agreement with previous works that found $\xoffnorm$ to increase after mergers \citep[e.g.,][]{power2012}.

\subsubsection{Velocity offset response}
\label{sec:voff_response}

The offset between the velocities of the centre of mass and the most bound particle, normalised by the virial circular velocity, $\voffnorm$, is closely related to $\xoffnorm$.
Like $\xoffnorm$, $\voffnorm$ slowly decreases for the random sample with the secular relaxation process.
For the merger sample, both $\voff$ and $\xoff$ increase during the inflow of mass, because the contribution of the secondary halo to the centre of mass continually increases in this process.
However, the two properties start to evolve differently after the secondary halo becomes part of the main halo.
If we approximately treat the two merging haloes as a system, higher $\voff$ corresponds to higher kinetic energy, and higher $\xoff$ corresponds to higher potential energy.
Without the dissipation due to the complicated interactions between the two haloes' particles, the total energy would be conserved, which would cause $\voff$ to decrease when $\xoff$ increases, and vice versa.
However, the peak in $\voffnorm$ occurs at $\Tmerger=0.28$, slightly earlier than the local minimum of $\xoffnorm$, because some of the energy is lost in dissipation and the total energy decreases.
For the same reason, $\voffnorm$ does not experience a second period of increase while $\xoffnorm$ decreases after the first apocentre, and both properties decrease while the secondary halo is disrupted.
We mark an elbow-like feature at $\Tmerger=0.68$, which approximately separates an earlier, faster phase of decrease from a later, slower phase of decrease in $\voffnorm$.
This time is slightly earlier than the apocentre, and coincides with the second local maximum in $\xoffnorm$.
We hypothesise that after the apocentric passage, the mass that remains bound to the secondary halo moves towards the centre of the main halo, and some of the potential energy is again transformed into kinetic energy, which slows down the decrease in $\voffnorm$.

\subsubsection{Orientation response}
\label{sec:orientation_response}

We track the evolution of the halo's orientation through the direction of the major axis with respect to its direction at the time of merger.
At $\Tmerger=0$, $\cos\theta=1$ by construction, and the alignment gradually decreases both backward and forward in time from this point, as the orientation changes.
The random sample exhibits a significant level of alignment over time, showing that the median evolution of the halo orientation is relatively mild.
Compared to the random sample, the orientation changes more rapidly for the merger sample, and shows a larger scatter.
There are two likely causes for this.
First, the merger temporarily elongates the main halo, and the direction of elongation affects the direction of the major axis.
Second, the merger brings in additional angular momentum, which enhances the rotation of the halo, and any rotation that is perpendicular to the major axis would lead to a change in the major axis direction.
Both these effects lead to stronger and more diverse evolution tracks of halo orientations for the merger sample.
The response of the merger sample suggests that the expedited change in the orientation is most significant immediately after the time of merger.
Without further quantitative investigation, we mark an elbow-like feature at $\Tmerger=0.32$ to guide the eye.

\subsection{Response dependence on cosmic time, merger ratio, and mass}
\label{sec:zrm_dep}

Having examined the qualitative response of properties to mergers, we now investigate some of the factors that quantitatively affect the merger response.
For an individual halo, there are many factors that may affect the evolution following a merger: the masses and internal properties of both haloes, the mass ratio of the merger, the relative velocity and impact factor of the merger, the preceding and following mass assembly history of the main halo, etc.
However, the detailed response to individual mergers is beyond the scope of this work.
We focus on the statistical influence on property responses of the three features described in \autoref{sec:merger_selection} --- the time of merger, $\amerger$, the merger ratio, $M_2/M_1$, and the previous halo mass, $M_{-1}$.

\begin{figure*}
    \centering
    \includegraphics[scale=0.5]{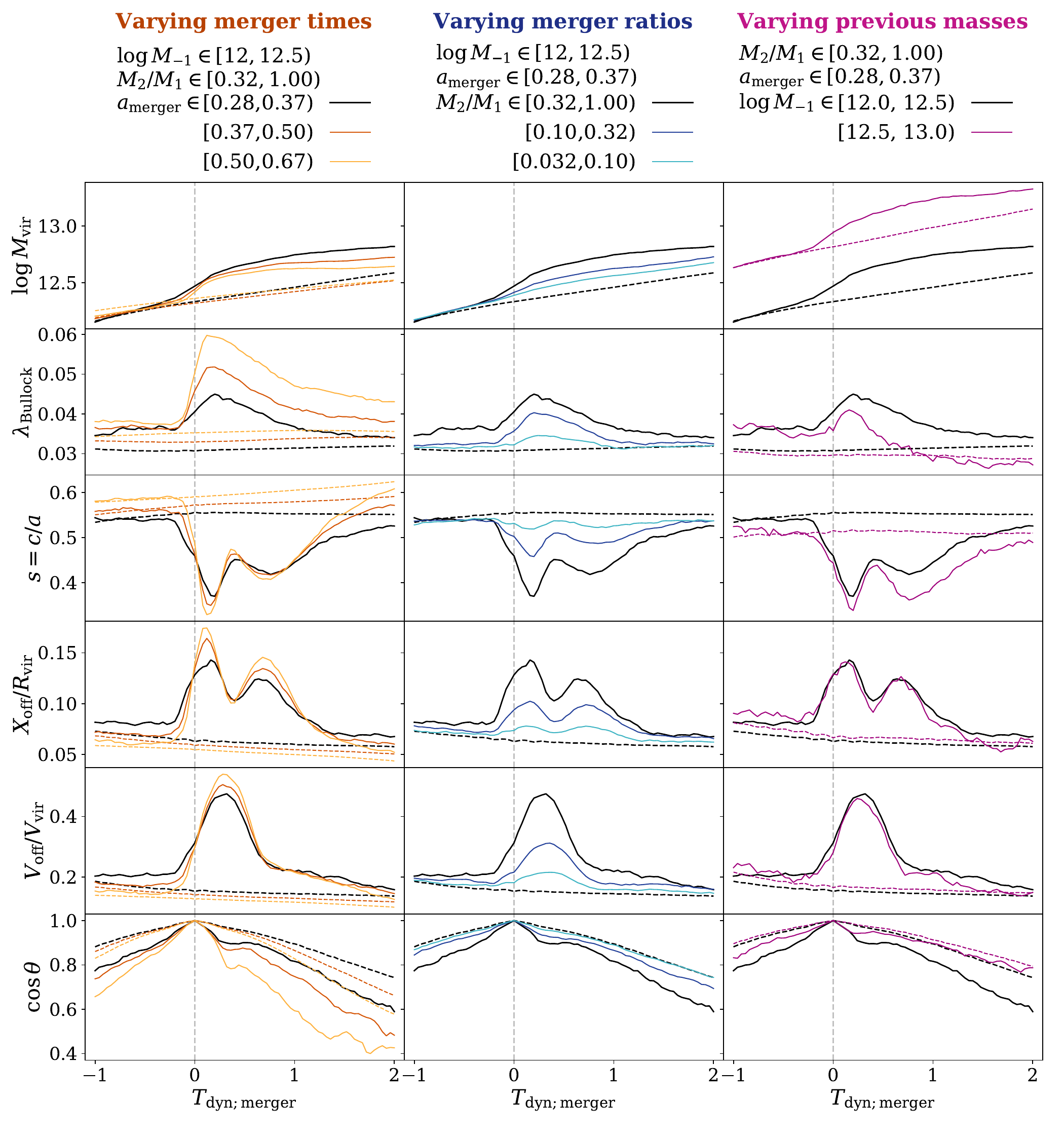}
    \caption{Comparison of stacked merger response between samples with different times of merger, merger ratios, and previous masses.
    Each column has two fixed features and one varying feature, which is labelled at the top.
    Each row shows the evolution of a different halo property.
    In each column we use solid lines of different colours to plot the property evolution of different merger samples, and the selection criteria for the samples are laid out at the top of the column.
    The dashed lines show the evolution of corresponding random samples in the same colours, except in the middle column, where the three merger samples correspond to the same random sample.
    The vertical dashed line in each panel indicates the time of merger.}
    \label{fig:response_dependences}
\end{figure*}

In \autoref{fig:response_dependences}, we demonstrate the influence of each feature, by comparing between the median property responses of different merger samples.
In each column, we vary one feature, and the detailed selection criteria are labelled at the top of the column.
One of the samples is kept unchanged in all columns and shown in black, with
\begin{align*}
    0.28\leq\amerger<0.37;\\
    0.32\leq M_2/M_1<1.00;\\
    12.0\leq\log M_{-1}<12.5.
\end{align*}
We make this choice to focus on major mergers, and also in consideration of the available sample sizes.
We test that comparisons between alternative samples yield qualitatively similar results.
For each merger sample, we also show the evolution of the corresponding random sample in the same colour with dashed curves.
In the middle column, the three merger samples correspond to the same random sample, which is shown as black dashed curves.
We discuss the potential caveats that arise from our sample selection in \autoref{sec:mthr_bias}.

An immediate observation we can make from \autoref{fig:response_dependences} is that, for each property, the shapes of the stacked responses are qualitatively similar with similar dynamical timescales, across different merger samples.
This confirms that the qualitative universality in the merger response of halo concentration that we found in W20 is applicable to the anisotropy properties studied in this work.
We may reasonably expect that the same is true for any halo property that is sensitive to the dynamical processes in halo mergers, because of the qualitative universality of merger dynamics.

\subsubsection{Varying merger times}
\label{sec:vary_amerger}

In the left column of \autoref{fig:response_dependences}, we compare merger samples selected at different times, with the same ranges of previous mass and merger ratio, to study the statistical dependence of merger responses on the cosmic time.

First, we compare between the random samples (dashed curves) that correspond to the merger samples, that have the same previous masses at different times.
We find systematic differences in all of the properties we study.
Random samples selected at later times have slightly lower mass growth rates, consistent with the prediction of $\Lambda$CDM cosmology that halo mass assembly slows down with time (although we note that our mass selection means that these random haloes do not have the same accretion histories as a typical $\Lambda$CDM halo with the same mass, see Section \ref{sec:random_selection}.)
Later random samples also have higher spins, more spherical shapes, smaller position and velocity offsets.
These findings suggest that the scaling relations with mass evolves upward for $\spinB$ or $c/a$, and downward for $\xoffnorm$ and $\voffnorm$ with time. 
We also find that $\cos\theta$ evolves faster for the later random samples.
At the same mass, haloes at earlier times correspond to rarer density peaks and are usually more dominant in their neighbourhoods.
They are therefore less likely to experience changes in orientation.

We now compare between the merger samples (solid curves).
The mass increase immediately around the time of merger is similar between the samples with different $\amerger$.
However, earlier merger samples have higher mass growth rates over longer times, like the random samples.
We find that mergers that happen at later times cause stronger responses in all the anisotropy properties.
A possible cause of this systematic difference is that the earlier samples tend to experience more frequent mergers and accrete more mass.
If different mergers happen along different directions, the combined effect of multiple mergers will reduce the level of anisotropy,  and lead to weaker responses in the anisotropy properties.
In other words, the impact of the single merger on halo anisotropy will be diluted by other merger activity, and appear less important.
This is a unique feature of the anisotropy properties.
Similarly, more frequent mergers suppress the change in halo orientation caused by the single merger, and the earlier samples show less evolution in $\cos\theta$. However, determining the strength and importance of this effect and disentangling it from, e.g., the difference in orbital isotropy of infalling subhaloes for hosts with different beak heights \citep{jiang_2015}, would require further testing.

\subsubsection{Varying merger ratios}
\label{sec:vary_ratio}

In the middle column of \autoref{fig:response_dependences}, we compare merger samples with the same ranges of merger time and previous mass, but different merger ratios.
These merger samples share the same corresponding random sample.
We find that mergers with higher ratios, i.e., more major mergers, induce more significant responses in properties.
More specifically, while the qualitative shapes of the responses are similar across the samples, the samples with higher merger ratios show higher amplitudes in their deviations from the secular evolution represented by the random sample.
This is consistent with our expectation that more major mergers lead to violent relaxation, and are able to cause more dramatic changes in the main halo.

We notice a small offset in the locations of the peaks and troughs between the samples with different merger ratios.
Specifically, the same oscillatory features tend to occur slightly later for the samples with lower merger ratios.
In \autoref{sec:diff_orbit_time}, we will show that less major mergers tend to have longer timescales, and it is reasonable to expect that longer timescales in the orbits are reflected in property responses, consistent with the offset we observe here.

\subsubsection{Varying previous masses}
\label{sec:vary_prevmass}

In the right column of \autoref{fig:response_dependences}, we compare merger samples with the same ranges of merger ratio and time of merger, but different previous masses.
Limited by sample sizes, we only compare between two bins of previous masses.

For both the random and merger samples, we do not find a clear dependence of the mass evolution on the previous mass, and the mass growth appears self-similar in different mass regimes (see also \autoref{fig:mthr_bias}).
The halo orientation evolves more strongly for less massive haloes, because they are less likely to be the dominant haloes in their own environments, and thus more sensitive to the surrounding gravitational interactions.

We now look at the internal anisotropy properties.
We find that $\spinB$ and $c/a$ are lower, and $\xoffnorm$ and $\voffnorm$ are higher for the more massive random sample.
This suggests that at the same cosmic time, $\spinB$ and $c/a$ are negatively correlated with halo mass, whereas $\xoffnorm$ and $\voffnorm$ are positively correlated with mass.
For mergers, we find that the more massive sample has slightly weaker responses, but the evidence is inconclusive, because of the noise in the measurement.
We hypothesise that this is because the mass accretion for the more massive haloes is more isotropic \citep{jiang_2015}, which suppresses the responses of anisotropy properties.

\section{Discussions}
\label{sec:discussion}

In this work, we have developed a further understanding of the dynamical mechanism of the evolution of halo anisotropy properties.
There have been previous works that study the connection between merger activity and the evolution of halo properties \citep[e.g.,][]{maccio08,ludlow2012,bett2016,lee2018}.
These works studied various halo properties and pointed out that they are systematically biased for halo populations with recent mergers.
Our findings are in general agreement with the previous understanding, but we further provide physically motivated explanations of the measured biases.
We underline the importance of measuring time in units of dynamical times, which aligns the response features and clarifies the connection between property responses and the orbital behaviour of the merging haloes.

Our findings show that halo mergers induce non-monotonic responses in these properties, which reflect the change of dynamical state of the halo.
Past mergers introduce systematic biases in the measured anisotropy properties in a non-trivial way, depending on the interval between the time of merger and the time of measurement.
This is a potential source of scatter in the scaling relations of the properties with halo mass and their correlations with each other \citep[see, e.g.,][for studies on such relations]{maccio07,jeeson-daniel2011,degraaff2022}.
These relations can be used in modelling halo properties in observational data, and for generating mock catalogues of various halo properties from basic properties.
A possible follow-up work is to quantitatively investigate the contribution of different mergers to scatters in the relations.

We have developed a framework to comprehensively study the evolution of halo properties ensuing halo mergers, which can also be applied to galaxy properties and/or galaxy mergers.
As a next step, we can readily analyse the other halo and galaxy properties calculated in \citetinprep{Anbajagane}, for example, halo velocity dispersion, galaxy stellar mass and gas mass, etc.
This analysis will reveal how each property responds to mergers, and also provide insight into the co-evolution of galaxy properties and halo properties, which informs the statistical connection between galaxies and haloes \citep[e.g.,][]{wechsler_tinker18}.
Another possible test is a comparison of halo property responses between the baryonic run of the simulation suite, and the dark matter-only counterpart.
This analysis can serve as a test of the baryonic effects on halo properties in the simulation.

Merger rates depend on the environment density, which, coupled with the merger responses, may be a source to the secondary halo biases \citep[e.g.,][the dependence of halo clustering on halo properties beyond mass]{gao_etal05,li_etal08} related to the anisotropy properties.
It would be interesting to explore the contribution of mergers to secondary halo biases, and possibly secondary galaxy biases \citep[e.g.,][]{croton_etal07} as well, as halo mergers can lead to galaxy mergers.

Findings from merger response studies may inform the design of observable proxies for the dynamical state of haloes.
For example, the X-ray gas morphology in galaxy clusters \citep[e.g.,][]{maurogordato2008,nurgaliev2017}, as quantitative merger response modelling may allow for for more physically grounded definitions of cluster ``relaxation.'' 
The fact that different halo properties exhibit different but related response timescales implies that deviations from mean scaling relations will be correlated across different properties, a fact which could be leveraged to reduce the scatter in masses inferred from these scaling relations and to produce more restrictive priors on the merger history of individual haloes.

Finally, our analyses are performed using IllustrisTNG data, and inherit all the modelling assumptions, as well as numerical issues, of the simulations.
We are also limited by the sample size available from the simulation volume.
The samples we use are based on the friends-of-friends, \textsc{Subfind} and \textsc{SubLink} algorithms, which introduce additional model dependence.
We do not expect our qualitative findings to be impacted by these choices.
However, future work with simulations that have larger volumes and higher resolutions, and alternative object finders and tree builders, may improve the quantitative significance of the results and enable more detailed case studies of merger events.

\section{Conclusion}
\label{sec:conclusions}

In this work, we use simulation data from IllustrisTNG to study how halo mergers impact the evolution of halo anisotropy properties.
Anisotropy properties measure the directional dependence of the halo internal structure, and are important for understanding the dynamical state of a halo.
We summarise our work below:
\begin{itemize}

    \item We generate catalogues of independently measured properties for a select sample of dark matter haloes from IllustrisTNG.
    In this work, we focus on the anisotropy properties (see \autoref{sec:props}).
    The details of the property measurement algorithm will be described in \citetinprep{Anbajagane}.
    
    \item We use a new package, \textsc{TreeHacker} \citepinprep{Wang}, to construct assembly histories and merger trees at the halo level, from the subhalo-based trees provided by IllustrisTNG.
    From the halo-based trees, we identify merger events and compile samples with physically motivated selection criteria.
    See \autoref{sec:methods} for details.
    
    \item For a sample of mergers, we measure characteristic timescales for the important dynamical stages during the merger process.
    These include the centre crossing, the first pericentric passage, and the first apocentric passage (see \autoref{fig:orbit_radius}).
    
    \item We examine the stacked evolution of the anisotropy properties during mergers and show that there is a significant deviation from secular evolution, illustrating a ``merger response'' in each of the properties (see \autoref{fig:uncertainty}).
    
    \item We associate qualitative features of the merger responses of anisotropy properties with the dynamical processes in mergers. 
    We identify some characteristic times in the merger responses (see \autoref{fig:uncertainty}).
    
    \item We assess systematic dependences of merger responses on the time of merger, merger ratio, and the mass of the main halo prior to the merger event. 
    The merger response exhibits a level of qualitative universality across these dependences, with variations manifesting in normalisation (see \autoref{fig:response_dependences}). 

\end{itemize}

\section*{Acknowledgements}
We thank Andrew Hearin, Yao-Yuan Mao, Daisuke Nagai, Risa Wechsler, Andrew Zentner, and members of the Baryon Pasting Collaboration, for useful discussions.

KW acknowledges support from the Leinweber Postdoctoral Research Fellowship at the University of Michigan.
PM acknowledges support from the Kavli Institute for Particle Astrophysics and Cosmology.
DA is supported by NSF grant No. 2108168.
CA acknowledges support from the Leinweber Center for Theoretical Physics and DOE grant DE- SC009193.

The IllustrisTNG simulations were undertaken with compute time awarded by the Gauss Centre for Supercomputing (GCS) under GCS Large-Scale Projects GCS-ILLU and GCS-DWAR on the GCS share of the supercomputer Hazel Hen at the High Performance Computing Center Stuttgart (HLRS), as well as on the machines of the Max Planck Computing and Data Facility (MPCDF) in Garching, Germany.
This research also made use of the Great Lakes Clusters at the University of Michigan, and the JupyterLab workspace hosted by the TNG team.

This research made use of Python, along with many community-developed or maintained software packages, including
IPython \citep{ipython},
Jupyter (\https{jupyter.org}),
Matplotlib \citep{matplotlib},
NumPy \citep{numpy},
SciPy \citep{scipy},
Pandas \citep{pandas}
and Astropy \citep{astropy}.
This research made use of NASA's Astrophysics Data System for bibliographic information.

\section*{Data Availability}
The simulation underlying this article were accessed from publicly available sources: \url{https://www.tng-project.org/data/}.
The derived catalogues will be made public in the near future, and shared on reasonable request to the corresponding authors before then.
The additional derived data are available in the article.



\bibliographystyle{mnras}
\bibliography{main,software}

\begin{thebibliography}{}
\makeatletter
\relax
\def\mn@urlcharsother{\let\do\@makeother \do\$\do\&\do\#\do\^\do\_\do\%\do\~}
\def\mn@doi{\begingroup\mn@urlcharsother \@ifnextchar [ {\mn@doi@}
  {\mn@doi@[]}}
\def\mn@doi@[#1]#2{\def\@tempa{#1}\ifx\@tempa\@empty \href
  {http://dx.doi.org/#2} {doi:#2}\else \href {http://dx.doi.org/#2} {#1}\fi
  \endgroup}
\def\mn@eprint#1#2{\mn@eprint@#1:#2::\@nil}
\def\mn@eprint@arXiv#1{\href {http://arxiv.org/abs/#1} {{\tt arXiv:#1}}}
\def\mn@eprint@dblp#1{\href {http://dblp.uni-trier.de/rec/bibtex/#1.xml}
  {dblp:#1}}
\def\mn@eprint@#1:#2:#3:#4\@nil{\def\@tempa {#1}\def\@tempb {#2}\def\@tempc
  {#3}\ifx \@tempc \@empty \let \@tempc \@tempb \let \@tempb \@tempa \fi \ifx
  \@tempb \@empty \def\@tempb {arXiv}\fi \@ifundefined
  {mn@eprint@\@tempb}{\@tempb:\@tempc}{\expandafter \expandafter \csname
  mn@eprint@\@tempb\endcsname \expandafter{\@tempc}}}

\bibitem[\protect\citeauthoryear{{Astropy Collaboration} et~al.,}{{Astropy
  Collaboration} et~al.}{2013}]{astropy}
{Astropy Collaboration} et~al., 2013, \mn@doi [\aap]
  {10.1051/0004-6361/201322068}, \href
  {http://adsabs.harvard.edu/abs/2013A%26A...558A..33A} {558, A33}

\bibitem[\protect\citeauthoryear{{Bett} \& {Frenk}}{{Bett} \&
  {Frenk}}{2016}]{bett2016}
{Bett} P.~E.,  {Frenk} C.~S.,  2016, \mn@doi [\mnras] {10.1093/mnras/stw1395},
  \href {https://ui.adsabs.harvard.edu/abs/2016MNRAS.461.1338B} {461, 1338}

\bibitem[\protect\citeauthoryear{{Bett}, {Eke}, {Frenk}, {Jenkins}  \&
  {Okamoto}}{{Bett} et~al.}{2010}]{bett2010}
{Bett} P.,  {Eke} V.,  {Frenk} C.~S.,  {Jenkins} A.,   {Okamoto} T.,  2010,
  \mn@doi [\mnras] {10.1111/j.1365-2966.2010.16368.x}, \href
  {https://ui.adsabs.harvard.edu/abs/2010MNRAS.404.1137B} {404, 1137}

\bibitem[\protect\citeauthoryear{{Blumenthal}, {Faber}, {Primack}  \&
  {Rees}}{{Blumenthal} et~al.}{1984}]{blumenthal_etal84}
{Blumenthal} G.~R.,  {Faber} S.~M.,  {Primack} J.~R.,   {Rees} M.~J.,  1984,
  \mn@doi [\nat] {10.1038/311517a0}, \href
  {https://ui.adsabs.harvard.edu/abs/1984Natur.311..517B} {311, 517}

\bibitem[\protect\citeauthoryear{{Boylan-Kolchin}, {Springel}, {White},
  {Jenkins}  \& {Lemson}}{{Boylan-Kolchin}
  et~al.}{2009}]{boylan-kolchin2009_milleniumII}
{Boylan-Kolchin} M.,  {Springel} V.,  {White} S. D.~M.,  {Jenkins} A.,
  {Lemson} G.,  2009, \mn@doi [\mnras] {10.1111/j.1365-2966.2009.15191.x},
  \href {https://ui.adsabs.harvard.edu/abs/2009MNRAS.398.1150B} {398, 1150}

\bibitem[\protect\citeauthoryear{{Bryan} \& {Norman}}{{Bryan} \&
  {Norman}}{1998}]{bryan_norman98}
{Bryan} G.~L.,  {Norman} M.~L.,  1998, \mn@doi [\apj] {10.1086/305262}, \href
  {https://ui.adsabs.harvard.edu/abs/1998ApJ...495...80B} {495, 80}

\bibitem[\protect\citeauthoryear{{Bullock} \& {Boylan-Kolchin}}{{Bullock} \&
  {Boylan-Kolchin}}{2017}]{bullock2017}
{Bullock} J.~S.,  {Boylan-Kolchin} M.,  2017, \mn@doi [\araa]
  {10.1146/annurev-astro-091916-055313}, \href
  {https://ui.adsabs.harvard.edu/abs/2017ARA&A..55..343B} {55, 343}

\bibitem[\protect\citeauthoryear{{Bullock}, {Kolatt}, {Sigad}, {Somerville},
  {Kravtsov}, {Klypin}, {Primack}  \& {Dekel}}{{Bullock}
  et~al.}{2001}]{bullock01}
{Bullock} J.~S.,  {Kolatt} T.~S.,  {Sigad} Y.,  {Somerville} R.~S.,  {Kravtsov}
  A.~V.,  {Klypin} A.~A.,  {Primack} J.~R.,   {Dekel} A.,  2001, \mn@doi
  [\mnras] {10.1046/j.1365-8711.2001.04068.x}, \href
  {https://ui.adsabs.harvard.edu/abs/2001MNRAS.321..559B} {321, 559}

\bibitem[\protect\citeauthoryear{{Chen}, {Mo}, {Li}, {Wang}, {Yang}, {Zhang}
  \& {Wang}}{{Chen} et~al.}{2020}]{chen2020}
{Chen} Y.,  {Mo} H.~J.,  {Li} C.,  {Wang} H.,  {Yang} X.,  {Zhang} Y.,   {Wang}
  K.,  2020, \mn@doi [\apj] {10.3847/1538-4357/aba597}, \href
  {https://ui.adsabs.harvard.edu/abs/2020ApJ...899...81C} {899, 81}

\bibitem[\protect\citeauthoryear{{Corless} \& {King}}{{Corless} \&
  {King}}{2007}]{corless2007}
{Corless} V.~L.,  {King} L.~J.,  2007, \mn@doi [\mnras]
  {10.1111/j.1365-2966.2007.12018.x}, \href
  {https://ui.adsabs.harvard.edu/abs/2007MNRAS.380..149C} {380, 149}

\bibitem[\protect\citeauthoryear{{Croton}, {Gao}  \& {White}}{{Croton}
  et~al.}{2007}]{croton_etal07}
{Croton} D.~J.,  {Gao} L.,   {White} S.~D.~M.,  2007, \mn@doi [\mnras]
  {10.1111/j.1365-2966.2006.11230.x}, \href
  {http://adsabs.harvard.edu/abs/2007MNRAS.374.1303C} {374, 1303}

\bibitem[\protect\citeauthoryear{{Davis}, {Efstathiou}, {Frenk}  \&
  {White}}{{Davis} et~al.}{1985}]{davis85_fof}
{Davis} M.,  {Efstathiou} G.,  {Frenk} C.~S.,   {White} S.~D.~M.,  1985,
  \mn@doi [\apj] {10.1086/163168}, \href
  {https://ui.adsabs.harvard.edu/abs/1985ApJ...292..371D} {292, 371}

\bibitem[\protect\citeauthoryear{{De Lucia} \& {Blaizot}}{{De Lucia} \&
  {Blaizot}}{2007}]{de-lucia2007}
{De Lucia} G.,  {Blaizot} J.,  2007, \mn@doi [\mnras]
  {10.1111/j.1365-2966.2006.11287.x}, \href
  {https://ui.adsabs.harvard.edu/abs/2007MNRAS.375....2D} {375, 2}

\bibitem[\protect\citeauthoryear{{Debattista}, {Moore}, {Quinn}, {Kazantzidis},
  {Maas}, {Mayer}, {Read}  \& {Stadel}}{{Debattista}
  et~al.}{2008}]{debattista2008}
{Debattista} V.~P.,  {Moore} B.,  {Quinn} T.,  {Kazantzidis} S.,  {Maas} R.,
  {Mayer} L.,  {Read} J.,   {Stadel} J.,  2008, \mn@doi [\apj]
  {10.1086/587977}, \href
  {https://ui.adsabs.harvard.edu/abs/2008ApJ...681.1076D} {681, 1076}

\bibitem[\protect\citeauthoryear{{Diemer} \& {Kravtsov}}{{Diemer} \&
  {Kravtsov}}{2014}]{diemer_kravtsov14}
{Diemer} B.,  {Kravtsov} A.~V.,  2014, \mn@doi [\apj]
  {10.1088/0004-637X/789/1/1}, \href
  {http://adsabs.harvard.edu/abs/2014ApJ...789....1D} {789, 1}

\bibitem[\protect\citeauthoryear{{Diemer}, {Behroozi}  \& {Mansfield}}{{Diemer}
  et~al.}{2023}]{diemer_2023}
{Diemer} B.,  {Behroozi} P.,   {Mansfield} P.,  2023, \mn@doi [arXiv e-prints]
  {10.48550/arXiv.2305.00993}, \href
  {https://ui.adsabs.harvard.edu/abs/2023arXiv230500993D} {p. arXiv:2305.00993}

\bibitem[\protect\citeauthoryear{{Faltenbacher}, {Li}, {White}, {Jing},
  {Shu-DeMao}  \& {Wang}}{{Faltenbacher} et~al.}{2009}]{faltenbacher2009}
{Faltenbacher} A.,  {Li} C.,  {White} S. D.~M.,  {Jing} Y.-P.,  {Shu-DeMao}
  {Wang} J.,  2009, \mn@doi [Research in Astronomy and Astrophysics]
  {10.1088/1674-4527/9/1/004}, \href
  {https://ui.adsabs.harvard.edu/abs/2009RAA.....9...41F} {9, 41}

\bibitem[\protect\citeauthoryear{{Gao}, {Springel}  \& {White}}{{Gao}
  et~al.}{2005}]{gao_etal05}
{Gao} L.,  {Springel} V.,   {White} S. D.~M.,  2005, \mn@doi [\mnras]
  {10.1111/j.1745-3933.2005.00084.x}, \href
  {https://ui.adsabs.harvard.edu/abs/2005MNRAS.363L..66G} {363, L66}

\bibitem[\protect\citeauthoryear{{Hahn}, {Carollo}, {Porciani}  \&
  {Dekel}}{{Hahn} et~al.}{2007}]{hahn2007b}
{Hahn} O.,  {Carollo} C.~M.,  {Porciani} C.,   {Dekel} A.,  2007, \mn@doi
  [\mnras] {10.1111/j.1365-2966.2007.12249.x10.48550/arXiv.0704.2595}, \href
  {https://ui.adsabs.harvard.edu/abs/2007MNRAS.381...41H} {381, 41}

\bibitem[\protect\citeauthoryear{{Han}, {Cole}, {Frenk}  \& {Jing}}{{Han}
  et~al.}{2016}]{han_2016}
{Han} J.,  {Cole} S.,  {Frenk} C.~S.,   {Jing} Y.,  2016, \mn@doi [\mnras]
  {10.1093/mnras/stv2900}, \href
  {https://ui.adsabs.harvard.edu/abs/2016MNRAS.457.1208H} {457, 1208}

\bibitem[\protect\citeauthoryear{{Hetznecker} \& {Burkert}}{{Hetznecker} \&
  {Burkert}}{2006}]{hetznecker2006}
{Hetznecker} H.,  {Burkert} A.,  2006, \mn@doi [\mnras]
  {10.1111/j.1365-2966.2006.10616.x}, \href
  {https://ui.adsabs.harvard.edu/abs/2006MNRAS.370.1905H} {370, 1905}

\bibitem[\protect\citeauthoryear{Hunter}{Hunter}{2007}]{matplotlib}
Hunter J.~D.,  2007, \mn@doi [Computing in Science Engineering]
  {10.1109/MCSE.2007.55}, 9, 90

\bibitem[\protect\citeauthoryear{{Jeeson-Daniel}, {Dalla Vecchia}, {Haas}  \&
  {Schaye}}{{Jeeson-Daniel} et~al.}{2011}]{jeeson-daniel2011}
{Jeeson-Daniel} A.,  {Dalla Vecchia} C.,  {Haas} M.~R.,   {Schaye} J.,  2011,
  \mn@doi [\mnras] {10.1111/j.1745-3933.2011.01081.x}, \href
  {https://ui.adsabs.harvard.edu/abs/2011MNRAS.415L..69J} {415, L69}

\bibitem[\protect\citeauthoryear{{Jiang} \& {van den Bosch}}{{Jiang} \& {van
  den Bosch}}{2016}]{Jiang_vdB16}
{Jiang} F.,  {van den Bosch} F.~C.,  2016, \mn@doi [\mnras]
  {10.1093/mnras/stw439}, \href
  {https://ui.adsabs.harvard.edu/abs/2016MNRAS.458.2848J} {458, 2848}

\bibitem[\protect\citeauthoryear{{Jiang}, {Cole}, {Sawala}  \& {Frenk}}{{Jiang}
  et~al.}{2015}]{jiang_2015}
{Jiang} L.,  {Cole} S.,  {Sawala} T.,   {Frenk} C.~S.,  2015, \mn@doi [\mnras]
  {10.1093/mnras/stv053}, \href
  {https://ui.adsabs.harvard.edu/abs/2015MNRAS.448.1674J} {448, 1674}

\bibitem[\protect\citeauthoryear{Jones, Oliphant, Peterson  et~al.}{Jones
  et~al.}{2001}]{scipy}
Jones E.,  Oliphant T.,  Peterson P.,   et~al., 2001, {SciPy}: Open source
  scientific tools for {Python}, \url {http://www.scipy.org/}

\bibitem[\protect\citeauthoryear{{Kazantzidis}, {Zentner}  \&
  {Kravtsov}}{{Kazantzidis} et~al.}{2006}]{kazanrzidis_2006}
{Kazantzidis} S.,  {Zentner} A.~R.,   {Kravtsov} A.~V.,  2006, \mn@doi [\apj]
  {10.1086/500579}, \href
  {https://ui.adsabs.harvard.edu/abs/2006ApJ...641..647K} {641, 647}

\bibitem[\protect\citeauthoryear{{Klypin}, {Yepes}, {Gottl{\"o}ber}, {Prada}
  \& {He{\ss}}}{{Klypin} et~al.}{2016}]{klypin_etal16}
{Klypin} A.,  {Yepes} G.,  {Gottl{\"o}ber} S.,  {Prada} F.,   {He{\ss}} S.,
  2016, \mn@doi [\mnras] {10.1093/mnras/stw248}, \href
  {https://ui.adsabs.harvard.edu/abs/2016MNRAS.457.4340K} {457, 4340}

\bibitem[\protect\citeauthoryear{{Kravtsov}}{{Kravtsov}}{2013}]{kravtsov13}
{Kravtsov} A.~V.,  2013, \mn@doi [\apjl] {10.1088/2041-8205/764/2/L31}, \href
  {http://adsabs.harvard.edu/abs/2013ApJ...764L..31K} {764, L31}

\bibitem[\protect\citeauthoryear{{Lee}, {Primack}, {Behroozi},
  {Rodr{\'\i}guez-Puebla}, {Hellinger}  \& {Dekel}}{{Lee}
  et~al.}{2018}]{lee2018}
{Lee} C.~T.,  {Primack} J.~R.,  {Behroozi} P.,  {Rodr{\'\i}guez-Puebla} A.,
  {Hellinger} D.,   {Dekel} A.,  2018, \mn@doi [\mnras]
  {10.1093/mnras/sty2538}, \href
  {https://ui.adsabs.harvard.edu/abs/2018MNRAS.481.4038L} {481, 4038}

\bibitem[\protect\citeauthoryear{{Li}, {Mo}  \& {Gao}}{{Li}
  et~al.}{2008}]{li_etal08}
{Li} Y.,  {Mo} H.~J.,   {Gao} L.,  2008, \mn@doi [\mnras]
  {10.1111/j.1365-2966.2008.13667.x}, \href
  {https://ui.adsabs.harvard.edu/abs/2008MNRAS.389.1419L} {389, 1419}

\bibitem[\protect\citeauthoryear{{Ludlow}, {Navarro}, {Li}, {Angulo},
  {Boylan-Kolchin}  \& {Bett}}{{Ludlow} et~al.}{2012}]{ludlow2012}
{Ludlow} A.~D.,  {Navarro} J.~F.,  {Li} M.,  {Angulo} R.~E.,  {Boylan-Kolchin}
  M.,   {Bett} P.~E.,  2012, \mn@doi [\mnras]
  {10.1111/j.1365-2966.2012.21892.x}, \href
  {https://ui.adsabs.harvard.edu/abs/2012MNRAS.427.1322L} {427, 1322}

\bibitem[\protect\citeauthoryear{{Ludlow}, {Bose}, {Angulo}, {Wang},
  {Hellwing}, {Navarro}, {Cole}  \& {Frenk}}{{Ludlow}
  et~al.}{2016}]{ludlow2016}
{Ludlow} A.~D.,  {Bose} S.,  {Angulo} R.~E.,  {Wang} L.,  {Hellwing} W.~A.,
  {Navarro} J.~F.,  {Cole} S.,   {Frenk} C.~S.,  2016, \mn@doi [\mnras]
  {10.1093/mnras/stw1046}, \href
  {https://ui.adsabs.harvard.edu/abs/2016MNRAS.460.1214L} {460, 1214}

\bibitem[\protect\citeauthoryear{{Macci{\`o}}, {Dutton}, {van den Bosch},
  {Moore}, {Potter}  \& {Stadel}}{{Macci{\`o}} et~al.}{2007}]{maccio07}
{Macci{\`o}} A.~V.,  {Dutton} A.~A.,  {van den Bosch} F.~C.,  {Moore} B.,
  {Potter} D.,   {Stadel} J.,  2007, \mn@doi [\mnras]
  {10.1111/j.1365-2966.2007.11720.x}, \href
  {https://ui.adsabs.harvard.edu/abs/2007MNRAS.378...55M} {378, 55}

\bibitem[\protect\citeauthoryear{{Macci{\`o}}, {Dutton}  \& {van den
  Bosch}}{{Macci{\`o}} et~al.}{2008}]{maccio08}
{Macci{\`o}} A.~V.,  {Dutton} A.~A.,   {van den Bosch} F.~C.,  2008, \mn@doi
  [\mnras] {10.1111/j.1365-2966.2008.14029.x}, \href
  {https://ui.adsabs.harvard.edu/abs/2008MNRAS.391.1940M} {391, 1940}

\bibitem[\protect\citeauthoryear{{Mansfield} \& {Kravtsov}}{{Mansfield} \&
  {Kravtsov}}{2020}]{mansfield_kravtsov_2020}
{Mansfield} P.,  {Kravtsov} A.~V.,  2020, \mn@doi [\mnras]
  {10.1093/mnras/staa430}, \href
  {https://ui.adsabs.harvard.edu/abs/2020MNRAS.493.4763M} {493, 4763}

\bibitem[\protect\citeauthoryear{{Mansfield}, {Darragh-Ford}, {Wang}, {Nadler}
  \& {Wechsler}}{{Mansfield} et~al.}{2023}]{mansfield_2023}
{Mansfield} P.,  {Darragh-Ford} E.,  {Wang} Y.,  {Nadler} E.~O.,   {Wechsler}
  R.~H.,  2023, \mn@doi [arXiv e-prints] {10.48550/arXiv.2308.10926}, \href
  {https://ui.adsabs.harvard.edu/abs/2023arXiv230810926M} {p. arXiv:2308.10926}

\bibitem[\protect\citeauthoryear{{Marinacci} et~al.,}{{Marinacci}
  et~al.}{2018}]{marinacci2018_TNG}
{Marinacci} F.,  et~al., 2018, \mn@doi [\mnras] {10.1093/mnras/sty2206}, \href
  {https://ui.adsabs.harvard.edu/abs/2018MNRAS.480.5113M} {480, 5113}

\bibitem[\protect\citeauthoryear{{Maurogordato} et~al.,}{{Maurogordato}
  et~al.}{2008}]{maurogordato2008}
{Maurogordato} S.,  et~al., 2008, \mn@doi [\aap] {10.1051/0004-6361:20077614},
  \href {https://ui.adsabs.harvard.edu/abs/2008A&A...481..593M} {481, 593}

\bibitem[\protect\citeauthoryear{{McCarthy}, {Schaye}, {Bird}  \& {Le
  Brun}}{{McCarthy} et~al.}{2017}]{mccarthy2017_bahamas}
{McCarthy} I.~G.,  {Schaye} J.,  {Bird} S.,   {Le Brun} A. M.~C.,  2017,
  \mn@doi [\mnras] {10.1093/mnras/stw2792}, \href
  {https://ui.adsabs.harvard.edu/abs/2017MNRAS.465.2936M} {465, 2936}

\bibitem[\protect\citeauthoryear{McKinney}{McKinney}{2010}]{pandas}
McKinney W.,  2010, in van~der Walt S.,  Millman J.,  eds, Proceedings of the
  9th Python in Science Conference. pp 51 -- 56

\bibitem[\protect\citeauthoryear{{Mendoza}, {Mansfield}, {Wang}  \&
  {Avestruz}}{{Mendoza} et~al.}{2023}]{mendoza2023}
{Mendoza} I.,  {Mansfield} P.,  {Wang} K.,   {Avestruz} C.,  2023, \mn@doi
  [\mnras] {10.1093/mnras/stad1768}, \href
  {https://ui.adsabs.harvard.edu/abs/2023MNRAS.523.6386M} {523, 6386}

\bibitem[\protect\citeauthoryear{{Mo}, {van den Bosch}  \& {White}}{{Mo}
  et~al.}{2010}]{mo_vdb_white10}
{Mo} H.,  {van den Bosch} F.~C.,   {White} S.,  2010, {Galaxy Formation and
  Evolution}.
Cambridge University Press

\bibitem[\protect\citeauthoryear{{Moster}, {Somerville}, {Maulbetsch}, {van den
  Bosch}, {Macci{\`o}}, {Naab}  \& {Oser}}{{Moster} et~al.}{2010}]{moster10}
{Moster} B.~P.,  {Somerville} R.~S.,  {Maulbetsch} C.,  {van den Bosch} F.~C.,
  {Macci{\`o}} A.~V.,  {Naab} T.,   {Oser} L.,  2010, \mn@doi [\apj]
  {10.1088/0004-637X/710/2/903}, \href
  {http://adsabs.harvard.edu/abs/2010ApJ...710..903M} {710, 903}

\bibitem[\protect\citeauthoryear{{Naiman} et~al.,}{{Naiman}
  et~al.}{2018}]{naiman2018_TNG}
{Naiman} J.~P.,  et~al., 2018, \mn@doi [\mnras] {10.1093/mnras/sty618}, \href
  {https://ui.adsabs.harvard.edu/abs/2018MNRAS.477.1206N} {477, 1206}

\bibitem[\protect\citeauthoryear{{Navarro}, {Frenk}  \& {White}}{{Navarro}
  et~al.}{1997a}]{nfw97}
{Navarro} J.~F.,  {Frenk} C.~S.,   {White} S. D.~M.,  1997a, \mn@doi [\apj]
  {10.1086/304888}, \href
  {https://ui.adsabs.harvard.edu/abs/1997ApJ...490..493N} {490, 493}

\bibitem[\protect\citeauthoryear{{Navarro}, {Frenk}  \& {White}}{{Navarro}
  et~al.}{1997b}]{navarro_1997}
{Navarro} J.~F.,  {Frenk} C.~S.,   {White} S. D.~M.,  1997b, \mn@doi [\apj]
  {10.1086/304888}, \href
  {https://ui.adsabs.harvard.edu/abs/1997ApJ...490..493N} {490, 493}

\bibitem[\protect\citeauthoryear{{Nelson} et~al.,}{{Nelson}
  et~al.}{2018}]{nelson2018a_TNG}
{Nelson} D.,  et~al., 2018, \mn@doi [\mnras] {10.1093/mnras/stx3040}, \href
  {https://ui.adsabs.harvard.edu/abs/2018MNRAS.475..624N} {475, 624}

\bibitem[\protect\citeauthoryear{{Nelson} et~al.,}{{Nelson}
  et~al.}{2019}]{nelson2019a_TNG}
{Nelson} D.,  et~al., 2019, \mn@doi [Computational Astrophysics and Cosmology]
  {10.1186/s40668-019-0028-x}, \href
  {https://ui.adsabs.harvard.edu/abs/2019ComAC...6....2N} {6, 2}

\bibitem[\protect\citeauthoryear{{Neto} et~al.,}{{Neto} et~al.}{2007}]{neto07}
{Neto} A.~F.,  et~al., 2007, \mn@doi [\mnras]
  {10.1111/j.1365-2966.2007.12381.x}, \href
  {https://ui.adsabs.harvard.edu/abs/2007MNRAS.381.1450N} {381, 1450}

\bibitem[\protect\citeauthoryear{{Nurgaliev} et~al.,}{{Nurgaliev}
  et~al.}{2017}]{nurgaliev2017}
{Nurgaliev} D.,  et~al., 2017, \mn@doi [\apj] {10.3847/1538-4357/aa6db4}, \href
  {https://ui.adsabs.harvard.edu/abs/2017ApJ...841....5N} {841, 5}

\bibitem[\protect\citeauthoryear{{Patiri}, {Cuesta}, {Prada}, {Betancort-Rijo}
  \& {Klypin}}{{Patiri} et~al.}{2006}]{patiri2006}
{Patiri} S.~G.,  {Cuesta} A.~J.,  {Prada} F.,  {Betancort-Rijo} J.,   {Klypin}
  A.,  2006, \mn@doi [\apjl] {10.1086/510330}, \href
  {https://ui.adsabs.harvard.edu/abs/2006ApJ...652L..75P} {652, L75}

\bibitem[\protect\citeauthoryear{P\'erez \& Granger}{P\'erez \&
  Granger}{2007}]{ipython}
P\'erez F.,  Granger B.~E.,  2007, \mn@doi [Computing in Science Engineering]
  {10.1109/MCSE.2007.53}, 9, 21

\bibitem[\protect\citeauthoryear{{Pillepich} et~al.,}{{Pillepich}
  et~al.}{2018}]{pillepich2018b_TNG}
{Pillepich} A.,  et~al., 2018, \mn@doi [\mnras] {10.1093/mnras/stx3112}, \href
  {https://ui.adsabs.harvard.edu/abs/2018MNRAS.475..648P} {475, 648}

\bibitem[\protect\citeauthoryear{{Planck Collaboration} et~al.,}{{Planck
  Collaboration} et~al.}{2014}]{planck13}
{Planck Collaboration} et~al., 2014, \mn@doi [\aap]
  {10.1051/0004-6361/201321591}, \href
  {https://ui.adsabs.harvard.edu/abs/2014A&A...571A..16P} {571, A16}

\bibitem[\protect\citeauthoryear{{Planck Collaboration} et~al.,}{{Planck
  Collaboration} et~al.}{2016}]{planck2016}
{Planck Collaboration} et~al., 2016, \mn@doi [\aap]
  {10.1051/0004-6361/201525830}, \href
  {https://ui.adsabs.harvard.edu/abs/2016A&A...594A..13P} {594, A13}

\bibitem[\protect\citeauthoryear{{Planck Collaboration} et~al.,}{{Planck
  Collaboration} et~al.}{2018}]{Planck2018_cosmoparams}
{Planck Collaboration} et~al., 2018, arXiv e-prints, \href
  {https://ui.adsabs.harvard.edu/\#abs/2018arXiv180706209P} {p.
  arXiv:1807.06209}

\bibitem[\protect\citeauthoryear{{Power}, {Knebe}  \& {Knollmann}}{{Power}
  et~al.}{2012}]{power2012}
{Power} C.,  {Knebe} A.,   {Knollmann} S.~R.,  2012, \mn@doi [\mnras]
  {10.1111/j.1365-2966.2011.19820.x}, \href
  {https://ui.adsabs.harvard.edu/abs/2012MNRAS.419.1576P} {419, 1576}

\bibitem[\protect\citeauthoryear{{Rodriguez-Gomez} et~al.,}{{Rodriguez-Gomez}
  et~al.}{2015}]{rodriguez-gomez2015}
{Rodriguez-Gomez} V.,  et~al., 2015, \mn@doi [\mnras] {10.1093/mnras/stv264},
  \href {https://ui.adsabs.harvard.edu/abs/2015MNRAS.449...49R} {449, 49}

\bibitem[\protect\citeauthoryear{{Rodr{\'\i}guez-Puebla}, {Behroozi},
  {Primack}, {Klypin}, {Lee}  \& {Hellinger}}{{Rodr{\'\i}guez-Puebla}
  et~al.}{2016}]{bolplanck2016}
{Rodr{\'\i}guez-Puebla} A.,  {Behroozi} P.,  {Primack} J.,  {Klypin} A.,  {Lee}
  C.,   {Hellinger} D.,  2016, \mn@doi [\mnras] {10.1093/mnras/stw1705}, \href
  {https://ui.adsabs.harvard.edu/abs/2016MNRAS.462..893R} {462, 893}

\bibitem[\protect\citeauthoryear{{Schaye} et~al.,}{{Schaye}
  et~al.}{2015}]{schaye2015_eagle}
{Schaye} J.,  et~al., 2015, \mn@doi [\mnras] {10.1093/mnras/stu2058}, \href
  {https://ui.adsabs.harvard.edu/abs/2015MNRAS.446..521S} {446, 521}

\bibitem[\protect\citeauthoryear{Springel}{Springel}{2010}]{springel_2010}
Springel V.,  2010, \mn@doi [Monthly Notices of the Royal Astronomical Society]
  {10.1111/j.1365-2966.2009.15715.x}, 401, 791

\bibitem[\protect\citeauthoryear{{Springel}, {White}, {Tormen}  \&
  {Kauffmann}}{{Springel} et~al.}{2001}]{subfind}
{Springel} V.,  {White} S. D.~M.,  {Tormen} G.,   {Kauffmann} G.,  2001,
  \mn@doi [\mnras] {10.1046/j.1365-8711.2001.04912.x}, \href
  {https://ui.adsabs.harvard.edu/abs/2001MNRAS.328..726S} {328, 726}

\bibitem[\protect\citeauthoryear{{Springel} et~al.,}{{Springel}
  et~al.}{2018}]{springel2018_TNG}
{Springel} V.,  et~al., 2018, \mn@doi [\mnras] {10.1093/mnras/stx3304}, \href
  {https://ui.adsabs.harvard.edu/abs/2018MNRAS.475..676S} {475, 676}

\bibitem[\protect\citeauthoryear{{Umetsu} et~al.,}{{Umetsu}
  et~al.}{2014}]{umetsu2014}
{Umetsu} K.,  et~al., 2014, \mn@doi [\apj] {10.1088/0004-637X/795/2/163}, \href
  {https://ui.adsabs.harvard.edu/abs/2014ApJ...795..163U} {795, 163}

\bibitem[\protect\citeauthoryear{{Vera-Ciro}, {Sales}, {Helmi}, {Frenk},
  {Navarro}, {Springel}, {Vogelsberger}  \& {White}}{{Vera-Ciro}
  et~al.}{2011}]{vera-ciro2011}
{Vera-Ciro} C.~A.,  {Sales} L.~V.,  {Helmi} A.,  {Frenk} C.~S.,  {Navarro}
  J.~F.,  {Springel} V.,  {Vogelsberger} M.,   {White} S. D.~M.,  2011, \mn@doi
  [\mnras] {10.1111/j.1365-2966.2011.19134.x}, \href
  {https://ui.adsabs.harvard.edu/abs/2011MNRAS.416.1377V} {416, 1377}

\bibitem[\protect\citeauthoryear{{Viola} et~al.,}{{Viola}
  et~al.}{2015}]{viola2015}
{Viola} M.,  et~al., 2015, \mn@doi [\mnras] {10.1093/mnras/stv1447}, \href
  {https://ui.adsabs.harvard.edu/abs/2015MNRAS.452.3529V} {452, 3529}

\bibitem[\protect\citeauthoryear{{Vitvitska}, {Klypin}, {Kravtsov}, {Wechsler},
  {Primack}  \& {Bullock}}{{Vitvitska} et~al.}{2002}]{vitvitska_2002}
{Vitvitska} M.,  {Klypin} A.~A.,  {Kravtsov} A.~V.,  {Wechsler} R.~H.,
  {Primack} J.~R.,   {Bullock} J.~S.,  2002, \mn@doi [\apj] {10.1086/344361},
  \href {https://ui.adsabs.harvard.edu/abs/2002ApJ...581..799V} {581, 799}

\bibitem[\protect\citeauthoryear{{Wang}, {Mao}, {Zentner}, {Lange}, {van den
  Bosch}  \& {Wechsler}}{{Wang} et~al.}{2020}]{Wang2020concentration}
{Wang} K.,  {Mao} Y.-Y.,  {Zentner} A.~R.,  {Lange} J.~U.,  {van den Bosch}
  F.~C.,   {Wechsler} R.~H.,  2020, \mn@doi [\mnras] {10.1093/mnras/staa2733},
  \href {https://ui.adsabs.harvard.edu/abs/2020MNRAS.498.4450W} {498, 4450}

\bibitem[\protect\citeauthoryear{{Wechsler} \& {Tinker}}{{Wechsler} \&
  {Tinker}}{2018}]{wechsler_tinker18}
{Wechsler} R.~H.,  {Tinker} J.~L.,  2018, \mn@doi [\araa]
  {10.1146/annurev-astro-081817-051756}, \href
  {http://adsabs.harvard.edu/abs/2018ARA%26A..56..435W} {56, 435}

\bibitem[\protect\citeauthoryear{{Wechsler}, {Bullock}, {Primack}, {Kravtsov}
  \& {Dekel}}{{Wechsler} et~al.}{2002}]{wechsler02}
{Wechsler} R.~H.,  {Bullock} J.~S.,  {Primack} J.~R.,  {Kravtsov} A.~V.,
  {Dekel} A.,  2002, \mn@doi [\apj] {10.1086/338765}, \href
  {https://ui.adsabs.harvard.edu/abs/2002ApJ...568...52W} {568, 52}

\bibitem[\protect\citeauthoryear{{White} \& {Rees}}{{White} \&
  {Rees}}{1978}]{whiterees78}
{White} S.~D.~M.,  {Rees} M.~J.,  1978, \mn@doi [\mnras]
  {10.1093/mnras/183.3.341}, \href
  {https://ui.adsabs.harvard.edu/abs/1978MNRAS.183..341W} {183, 341}

\bibitem[\protect\citeauthoryear{{Wong} \& {Taylor}}{{Wong} \&
  {Taylor}}{2012}]{wong2012}
{Wong} A. W.~C.,  {Taylor} J.~E.,  2012, \mn@doi [\apj]
  {10.1088/0004-637X/757/1/102}, \href
  {https://ui.adsabs.harvard.edu/abs/2012ApJ...757..102W} {757, 102}

\bibitem[\protect\citeauthoryear{{Zemp}, {Gnedin}, {Gnedin}  \&
  {Kravtsov}}{{Zemp} et~al.}{2011}]{zemp2011}
{Zemp} M.,  {Gnedin} O.~Y.,  {Gnedin} N.~Y.,   {Kravtsov} A.~V.,  2011, \mn@doi
  [\apjs] {10.1088/0067-0049/197/2/30}, \href
  {https://ui.adsabs.harvard.edu/abs/2011ApJS..197...30Z} {197, 30}

\bibitem[\protect\citeauthoryear{{de Graaff}, {Trayford}, {Franx}, {Schaller},
  {Schaye}  \& {van der Wel}}{{de Graaff} et~al.}{2022}]{degraaff2022}
{de Graaff} A.,  {Trayford} J.,  {Franx} M.,  {Schaller} M.,  {Schaye} J.,
  {van der Wel} A.,  2022, \mn@doi [\mnras] {10.1093/mnras/stab3510}, \href
  {https://ui.adsabs.harvard.edu/abs/2022MNRAS.511.2544D} {511, 2544}

\bibitem[\protect\citeauthoryear{van~der Walt, Colbert  \& Varoquaux}{van~der
  Walt et~al.}{2011}]{numpy}
van~der Walt S.,  Colbert S.~C.,   Varoquaux G.,  2011, \mn@doi [Computing in
  Science Engineering] {10.1109/MCSE.2011.37}, 13, 22

\makeatother
\end{thebibliography}



\begin{appendices}

\counterwithin{figure}{section}
\counterwithin{table}{section}

\section{Orbit Timescales for Different Merger Ratios}
\label{sec:diff_orbit_time}

\begin{figure}
    \centering
    \includegraphics[scale=0.5]{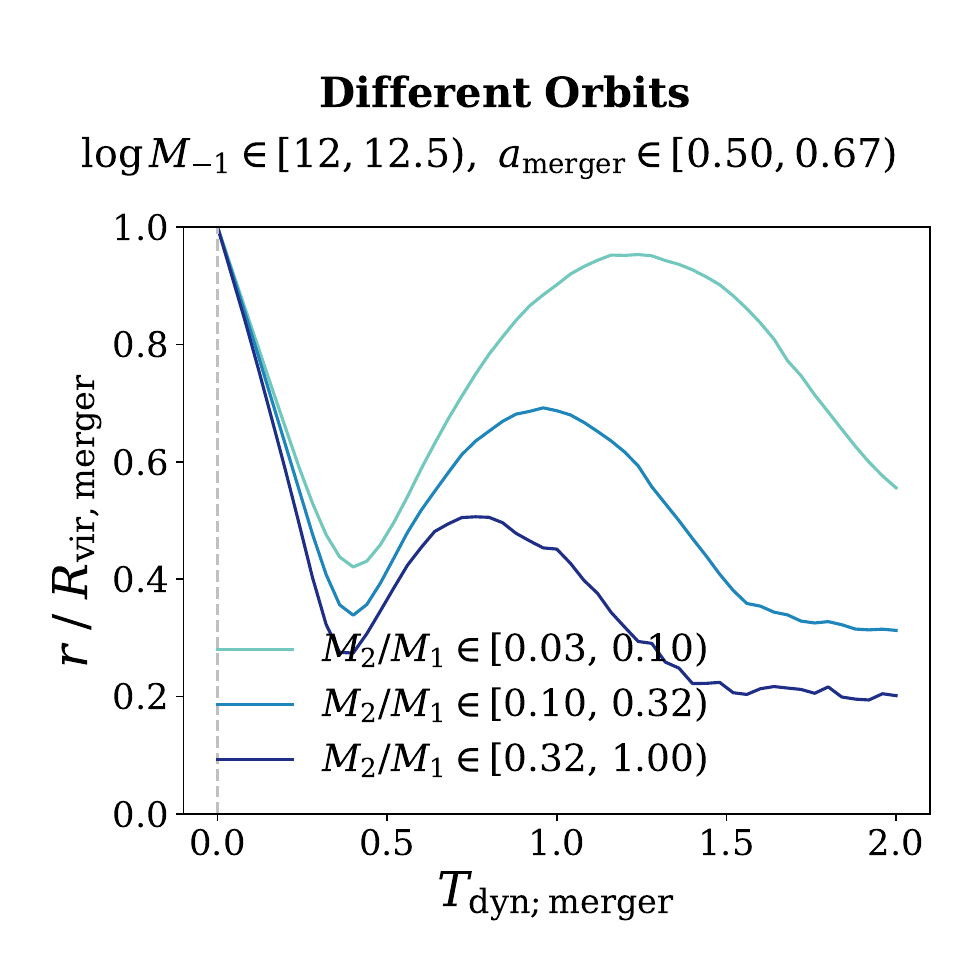}
    \caption{Orbits of secondary haloes for stacked merger samples with different merger ratios, as labelled in the figure.
    This figure is similar to \autoref{fig:orbit_radius}.
    For visual clarity, we only compare the median orbits and do not include the 16-84th percentile ranges.}
    \label{fig:orbit_ratio}
\end{figure}

In \autoref{sec:orb_tdyn}, we have measured the timescales of the orbits of the secondary haloes in mergers.
In this appendix, we look further into the dependence of orbit timescales on the merger ratio.
In addition to the sample with
\begin{align*}
    &0.5\leq\amerger<0.67,\\
    &0.32\leq M_2/M_1<1.00,\\
    &12\leq\log M_{-1}<12.5,
\end{align*}
studied in \autoref{sec:orb_tdyn}, we measure the median orbits for two other samples with the same $\amerger$ and $\log M_{-1}$, but with $0.10\leq M_2/M_1<0.32$ and $0.032\leq M_2/M_1<0.10$ respectively.
We show the results in \autoref{fig:orbit_ratio}.
Median orbits for the different merger samples are plotted in different colours, as labelled in the figure.
We find that the samples of less major mergers tend to orbit at larger radii and have longer timescales, in agreement with the temporal offsets in property responses found in \autoref{sec:vary_ratio}.
This is consistent with our expectations, because secondary haloes with smaller masses relative to the main halo experience less dynamical friction, and are less effective in transferring orbital energy to the main halo.
We caution that these results are also subject to the survivor bias discussed in \autoref{sec:orb_tdyn}, and may overestimate the timescales for the pericentric and apocentric passages.

\section{Test of Selection Bias from Mass Threshold}
\label{sec:mthr_bias}

The samples used in this work are based on the halo sample that includes the 10000 most massive halo in each snapshot of the TNG300-1 simulation box (see \autoref{sec:halo_samp}).
We choose this base halo sample to represent an approximately consistent population of haloes throughout cosmic time.
However, this algorithm leads to a moving mass threshold, which can cause a selection bias for our results in \autoref{sec:zrm_dep}.
For merger and random samples with previous masses below the moving mass threshold, higher mass growth rates are favoured in the following evolution, in order for the haloes to reach the threshold at later times and be included in the sample.

In this appendix, we use alternatively selected samples without the selection bias to test its impact on our results.
We lower the mass threshold of the base halo sample to $10^{11}\Msunh$, construct assembly histories, and select test merger and random samples for this base sample.
Unfortunately, it is computationally infeasible to perform the full analysis on this larger test sample, due to the runtime needed for calculating halo properties.
We limit our test to the mass evolution, using the virial masses provided in the TNG300-1 catalogues.

\begin{figure*}
    \centering
    \includegraphics[scale=0.55, trim = 0 0.7 2cm 0.7cm, clip]{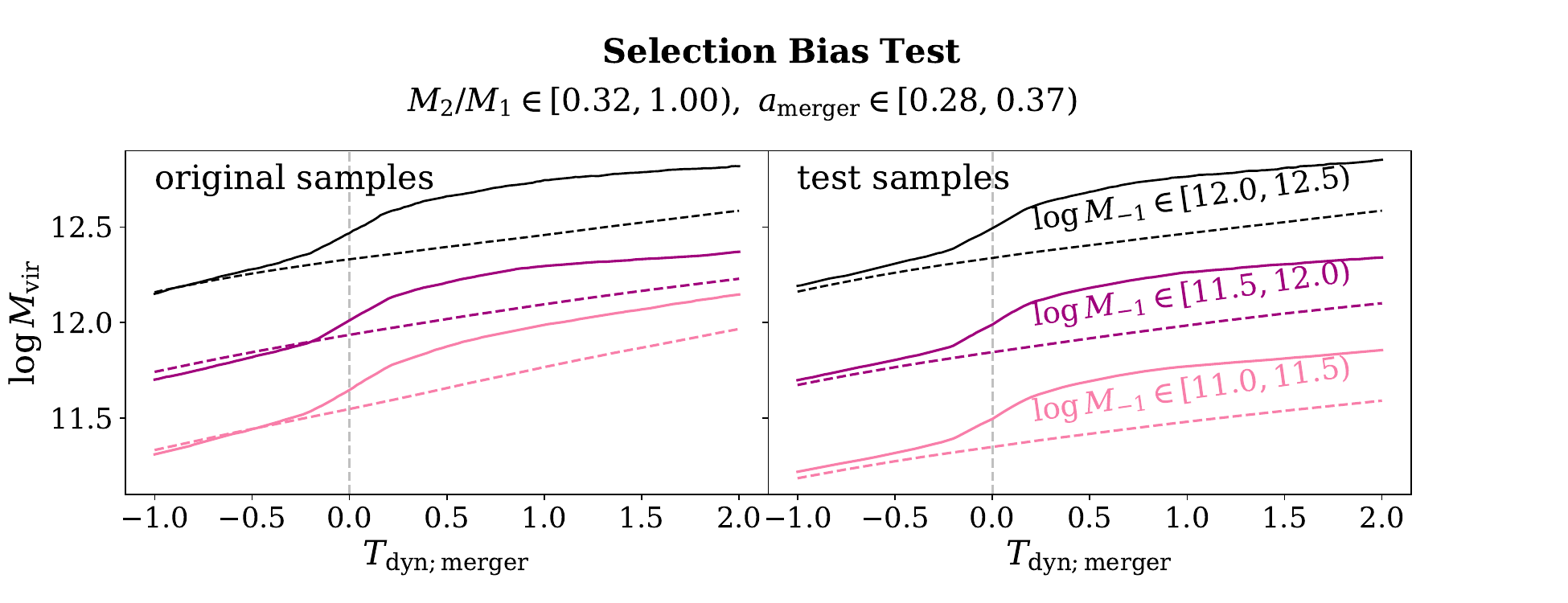}
    \caption{Test of selection bias from halo mass threshold.
    Each panel is similar to the top right panel of \autoref{fig:response_dependences}, but with different bins of previous masses, as labelled at the top of the figure and in the right panel.
    The comparison between the original and test samples shows that the selection bias introduces an artificial dependence of mass evolution on the previous mass.}
    \label{fig:mthr_bias}
\end{figure*}

For both the original samples used in the main text and the test samples defined above, we reproduce the analysis in the top right panel of \autoref{fig:response_dependences} with three lower bins of previous mass, $11\leq\log M_{-1}<11.5$, $11.5\leq\log M_{-1}<12$, and $12\leq\log M_{-1}<12.5$.
The other sample criteria are unchanged: $0.28\leq\amerger<0.37$ and $0.32\leq M_2/M_1<1.00$.
We show the results in \autoref{fig:mthr_bias}.
We find that for the test samples in the right panel, both the secular evolution and the merger response appear self-similar for different masses.
In comparison, for the original samples, at lower $M_{-1}$, the later mass growth rates of both the secular evolution and the merger response are artificially increased, as we expect.
Also, the median previous mass in each bin is higher than for the test sample, because haloes that have higher masses at earlier times are more likely to to have later masses above the moving mass threshold, and are favoured by the selection algorithm.

The highest bin with $12\leq\log M_{-1}<12.5$ is also used in the main text, for which the original and test results are largely consistent.
We show with this test that for $0.28\leq\amerger<0.37$, this mass is sufficiently high, and our results in \autoref{sec:vary_prevmass}, with $12\leq\log M_{-1}<12.5$ and $12.5\leq\log M_{-1}<13$, are robust to the selection bias.
On the other hand, the mass threshold is higher at later times, and the samples with $12\leq\log M_{-1}<12.5$ and $0.37\leq\amerger<0.50$ or $0.50\leq\amerger<0.67$ are more susceptible to the bias.
However, we argue that our qualitative conclusions in \autoref{sec:vary_amerger} are not affected, because the bias does not eliminate the expected qualitative dependence of the mass growth rate on the time of merger in the top left panel of \autoref{fig:response_dependences}.

\end{appendices}


\bsp	
\label{lastpage}
\end{document}